\documentclass[journal]{IEEEtran}
\usepackage{graphicx}
\usepackage{epstopdf}
\usepackage{bm}
\usepackage{cite}
\usepackage{amsmath,amssymb,amsfonts}
\usepackage{algorithm}
\usepackage{algorithmic}
\usepackage{graphicx}
\usepackage{textcomp}
\usepackage{xcolor}
\usepackage{epstopdf}
\usepackage{verbatim}
\usepackage{wasysym}
\usepackage{caption}
\usepackage{float}
\usepackage{subcaption}
\newtheorem{proposition}{Proposition}
\newtheorem{corollary}{Corollary}
\newtheorem{lemma}{Lemma}
\newenvironment{proof}{{\indent\it Proof:\ }}{\hfill $\blacksquare$\par}
\newenvironment{proof1}{{\indent\it Proof:\ }}{}
\usepackage{color}

\begin{document}
	\captionsetup[figure]{name={Fig.},labelsep=period,singlelinecheck=off}

	\title{Timeliness of Information for Computation-intensive Status Updates in  Task-oriented Communications}

	\author{\large Xiaoqi Qin,~\IEEEmembership{Member,~IEEE,}
		Yanlin Li,~\IEEEmembership{Student Member,~IEEE,} Xianxin Song,~\IEEEmembership{Student Member,~IEEE,}, \\Nan Ma,~\IEEEmembership{Member,~IEEE,} Chuan Huang,~\IEEEmembership{Member,~IEEE,} and Ping Zhang,~\IEEEmembership{Fellow,~IEEE}\\
		
		\thanks{X. Qin, Y. Li, N. Ma (\textit{corresponding author}) and P. Zhang are with the State Key Laboratory of Networking and Switching Technology, Beijing University of Posts and Telecommunications, Beijing, 100876, China, (e-mails: xiaoqiqin@bupt.edu.cn; yanlin@bupt.edu.cn; manan@bupt.edu.cn; pzhang@bupt.edu.cn).}
		\thanks{X. Song and C. Huang are with the School of Science and Engineering (SSE) and Future Network of Intelligence Institute (FNII), the Chinese University of Hong Kong, Shenzhen 518172, China, (emails: xianxinsong@cuhk.edu.cn; huangchuan@cuhk.edu.cn).}
	}
	
	\maketitle

\begin{abstract}
Moving beyond just interconnected devices, 
the increasing interplay between communication and computation has fed the vision of real-time networked control systems.
To obtain timely situational awareness,
IoT devices continuously sample computation-intensive status updates, generate perception tasks
and offload them to edge servers for processing.
In this sense,
the timeliness of information is considered as one major contextual attribute of status updates.
In this paper, 
we derive the closed-form expressions of timeliness of information for computation offloading at both edge tier and fog tier,
where two-stage tandem queues are exploited to abstract the transmission and computation process. 
Moreover, 
we exploit the statistical structure of Gauss-Markov process,
which is widely adopted to model temporal dynamics of system states, 
and derive the closed-form expression for process-related timeliness of information. 
The obtained analytical formulas explicitly characterize the dependency among task generation, transmission and execution, 
which can serve as objective functions for system optimization. 
Based on the theoretical results, 
we formulate a computation offloading optimization problem at edge tier, 
where the timeliness of status updates is minimized among multiple devices by joint optimization of task generation, bandwidth allocation, and computation resource allocation. 
An iterative solution procedure is proposed to solve the formulated problem. 
Numerical results reveal the intertwined relationship among transmission and computation stages, and verify the necessity of factoring in the task generation process for computation offloading strategy design.
\end{abstract}
	
	
	\IEEEpeerreviewmaketitle

	\section{Introduction}
	\label{sec:I:introduction}
	
Ushering in a new era of connected intelligence,
today's communications system is evolving towards a communicate-and-compute system,
where the scattered computing resources at multiple tiers of network function as distributed neurons that links the physical and cyber worlds,
propelling the expanding of emerging mission-critical interactive systems, exemplified by tactile Internet, autonomous networked control systems, and remote healthcare\cite{yang-natureelectronics}.
This new paradigm
involves information continuously flows around a control loop between physical devices and computation servers at network edge.
Let us consider,
for the sake of clarity,
a device generates perception tasks that contain computation-intensive status updates about time-varying physical phenomenon,
and offloads the tasks to edge server to extract the status information.
The casually extracted status updates are further employed for real-time reconstruction and estimation.
It necessitates extending the scope of system design objectives beyond error-free data transmission to encompass the context of information,
with focus on the \emph{usefulness} of information in relevant to the success of subsequent tasks \cite{ping-engineering}.

One major contextual attribute of extracted status updates is the end-to-end timing mismatch in relevance to the temporal dynamics at source device,
which is termed as \emph{timeliness of information} \cite{kountouris-semanticnetwork}.
It is directly related to transmission delay over wireless channels and computation delay at edge servers.
At system level,
the precious communication and computation resources should be carefully orchestrated to convey information that is valuable for subsequent tasks. 
Therefore, it calls for novel performance metrics focusing on task-oriented unification of information generation, transmission, and processing,
which have commonly been treated with separate design philosophy.
This problem, if left unsolved,
will impede the performance of real-time networked systems.

In this paper, 
we consider a multi-tier computing enabled networked control system.
IoT devices continuously generate perception tasks consisting of computation-intensive status updates, 
and offload them to network edge for processing. 
The timeliness of extracted status updates is jointly decided by task load and the queueing delay at both task offloading and processing stage.  
We aim to investigate the closed-form expressions of timeliness of information, 
which captures the intertwined relationship among task generation frequency, transmission rate, and computing speed,
so as can serve as a basis for system optimization to strike a balance between the timeliness of obtained status updates and system cost.

\subsection{Related Works}
\emph{ 1) Edge Computing:}
As a promising solution to boost the efficiency of networked information processing,
multi-tier computing network \cite{yang-iot, kato-multitier} has attracted extensive research interests.
To fully exploit the computation resources scattered at edge, fog and cloud,
efforts have been devoted to investigate computation offloading strategies to balance workload among different tiers,
subject to stringent delay requirements \cite{wang-queueedge,Yang20:IoT:OFDMA} or constrained energy budget \cite{Dinh:TCOM:partial-singleMEC,mao-dynamic}.
As for single-user scenarios,
the strategy design focuses on the trade-offs between offloading energy cost at user device and task completion delay \cite{Dinh:TCOM:partial-singleMEC,wang-queueedge}.
As for multi-user scenarios,
the computation offloading problem is more complicated due to resource contention,
where the allocation of bandwidth and computing resource are jointly optimized among users \cite{Yang20:IoT:OFDMA}.
The task offloading modes are classified into binary offloading and partial offloading. 
In case of binary offloading, 
the computational tasks are highly integrated, 
and thus are executed either locally at end device or remotely at MEC server \cite{mao-dynamic}.
In case of partial offloading,
the computation tasks are composed of several parallel components, 
and the system performance can be improved by optimizing the portion of data for local computing and offloading\cite{Dinh:TCOM:partial-singleMEC}. 
In this paper, 
we adopt the binary task offloading model.

In these existing works,
the offloading strategies focus on optimizing  \textit{one-shot} task execution.
However,
as for time-critical networked control scenarios considered in this paper,
status updates obtained by executing tasks are temporally correlated. 
Therefore, the consecutively generated tasks cannot be treated independently.
Different from existing work, 
we aim to analyze the timeliness of status updates obtained by \textit{constantly} processing computational task flows generated by IoT devices.
Due to stochastic task generation intervals and time-varying workloads at MEC server, 
a naive application of existing computation offloading strategies would lead to poor system performance. 
This motivates us to conceive this paper.

\emph{ 2) AoI for Tandem Queues:}
The concept of age of information (AoI)\cite{yates-aoi-jsac} extends delay metric to incorporate the information generation process.
The modeling of age has been evaluated with the aid of queuing theory,
where a source node generates data as a stochastic process and the transmission over wireless network is abstracted as
queues,
including M/M/1\cite{kaul-concept}, M/D/1\cite{inoue-md1}, and G/G/1/1\cite{soysal-g/g/1/1}.
In the event of queuing,
data packets start aging while waiting for transmission,
thus the \emph{zero-wait} policy is studied, where source node generates new packets only when the channel becomes idle \cite{sun-wait,yates-lazy}.
In this paper, the impact of zero-wait policy is investigated as a special case of our proposed model.


The adoption of age metric in edge computing system requires an extension of modeling choice from single transmission queue to two-stage tandem queues.
The ``compute-then-transmit'' mode is adopted in\cite{zou-management, xu-computingiot},
where sampled data is preprocessed locally before uploading to edge server.
The single user scenario is investigated in \cite{zou-management} using GI/M/1-M/GI/1 tandems with various queue management techniques,
where the closed-form expressions for averaged AoI and averaged peak AoI are derived.
Multiple users in the first queue is studied in \cite{xu-computingiot} as a priority M/G/1,
where the derived peak AoI expression is further employed for strategy design.
The ``transmit-then-compute'' mode for single user scenario is studied in \cite{chiariotti-peakdistribution,kuang-intensive},
where user offloads its tasks to edge server for execution. 
The full distribution of peak AoI is derived for both M/M/1-M/D/1 and M/M/1-M/M/1 tandems in \cite{chiariotti-peakdistribution},
which could be further employed to define reliability requirements.
In \cite{kuang-intensive},
The closed-form averaged AoI is derived for local computing, remote computing, and partial offloading mode under zero-wait policy. 

 The aforementioned works investigate linear age models that grows at unit rate.
 However, 
 it may not be the best choice for timeliness of extracted status updates investigated in this paper\cite{uysal-aoi+value}.
 To obtain a more meaningful representation, 
 we exploit the structure of monitored physical process to establish a process-related measure for timeliness of information,
 which can be derived by a combination of tandem queues with non-linear function of age. 
 The obtained closed-form expression reveals the real-time estimation error of current status at source, 
 which can be employed to guide the design of subsequent control task decisions\cite{niu-urgency}. 
 Moreover, 
 we consider resource contention among multiple devices for computation offloading at fog-tier, 
 which complicates the analysis of the second queue due to aggregation of tasks that experience stochastic services times at their first queue. 
 Such analysis is considered challenging even for single queue models \cite{yates-multisource}.

\subsection{Contributions}
The main contributions of this paper can be summarized as follows. 
\begin{itemize}
	\item We derive the closed-form expressions of timeliness of information for computation offloading at both edge and fog tier.
	The transmission and computation process are modeled as two queues in tandem.
	The obtained analytical formulas explicitly characterize the dependency among task generation, transmission and computation under stochastic resource availability, which can be employed as performance metrics for system optimization.  

	\item We take a closer look at the statistical property of monitored physical process, and establish a mathematical modeling framework for \emph{process-related} timeliness of information. 
	We are interested in the widely adopted Gauss-Markov process. 
	The mean squared estimation error is employed as a measure of timeliness of information, 
	which is modeled as a non-linear function of age.
	\item We also investigate \emph{zero-wait} task generation policy as special cases in all scenarios, and derive the corresponding closed-form expressions. The impact of zero-wait policy is numerically investigated. 
	\item Based on the analytic results, 
	we formulate a timeliness of information minimization problem for computation offloading at edge tier, 
	which jointly optimizes task generation, bandwidth allocation, and computation resource allocation. 
	Simulation results demonstrate the effectiveness of our proposed strategy. 
\end{itemize} 

The remainder of this paper is organized as follows.
In Sec. \ref{sec:II:systemModel},
we present the system architecture. 
In Sec. \ref{sec:II:ToIFramework},
we propose an analysis framework for timeliness of information. 
In Sec. \ref{sec:III:edgeTier} and Sec. \ref{sec:IV:fogTier},
we derive closed-form expressions for timeliness of information at edge and fog tier, respectively. 
In Sec. \ref{sec:V:caseStudy},
we formulate a computation offloading optimization problem at edge tier
and present simulation results. 
Sec. \ref{sec:VI:conclusion} concludes this paper.

\section{System Architecture}
\label{sec:II:systemModel}

\begin{figure}
	\raggedright{\includegraphics[scale = 0.285]{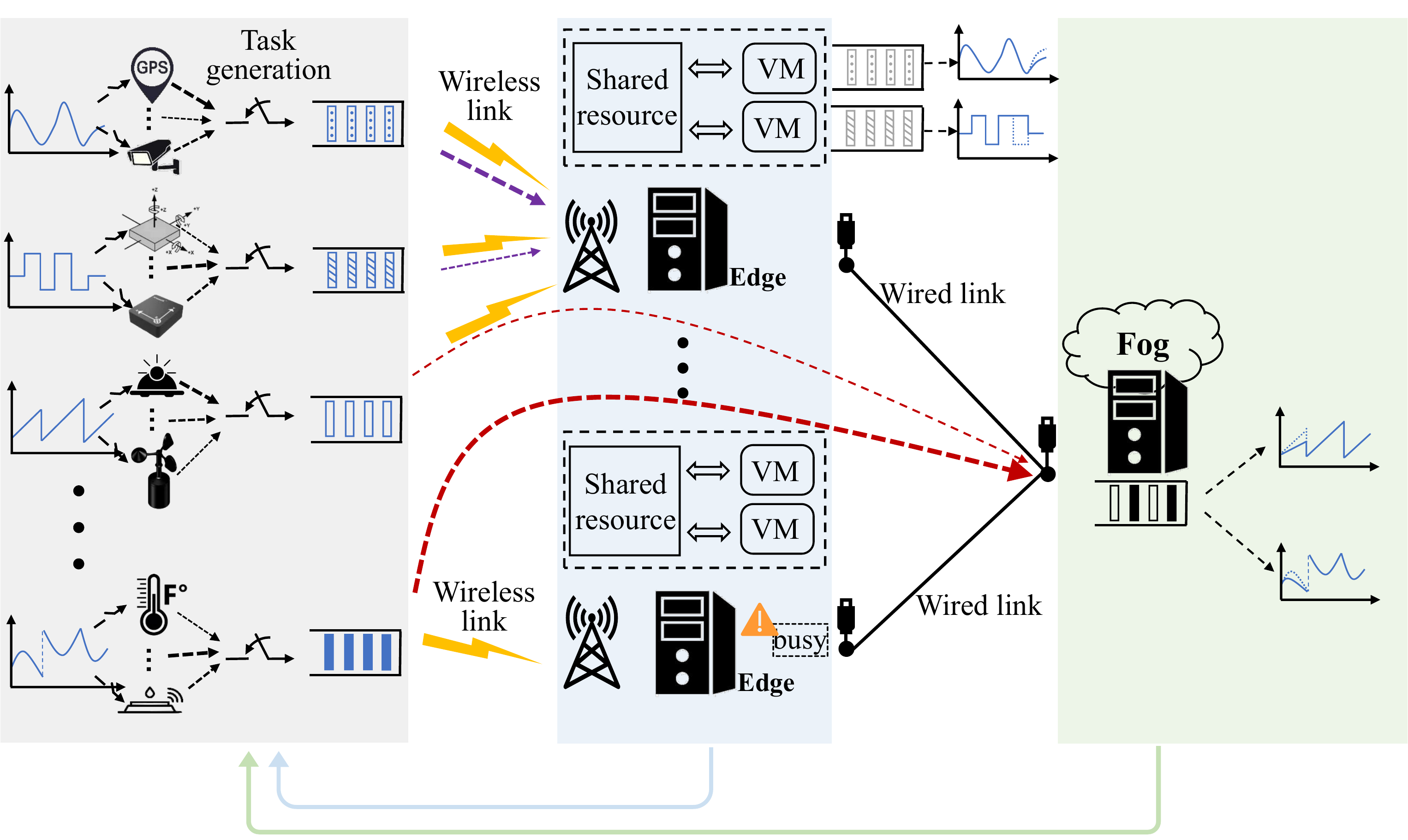}}
	\caption{An example of multi-tier computing enabled networked control system.}
	\label{system}
\end{figure}

We consider a multi-tier computing enabled networked control system as shown in Fig.~\ref{system},
which consists of a set of IoT devices, edge servers that are deployed in proximity with the devices,
and a fog server which provides computing service at regional level.
As shown in the figure,
each device continuously generates perception tasks and offloads them to edge server or fog server to extract status updates about its surroundings.
	
At edge tier,
edge servers are deployed in proximity to devices and interact with them via wireless links.
Assume that each edge server holds multiple virtual machines for parallel task execution among devices \cite{huang-parrallel}.	
Considering a realistic problem that the number of virtual machines at each server is usually constrained due to I/O interference,
the edge server can further outsource computation tasks to fog tier via wired links.
Therefore,
fog tier handles tasks relayed by multiple edge servers at regional level.
Due to the stochasticity of outsourced task arrivals,
we assume that they are sequentially executed at fog server with full computing power\cite{shen-cloudprovision},
so that computation resource is fully exploited to accelerate task execution\footnote{At fog tier, the tasks can also be processed in parallel with pre-assigned share of computing power. Here, we employ sequential task execution mode, which employs full computing power for each task, in order to provide a comprehensive performance analysis of various computing modes.}.

At each IoT device, 
we model the computation offloading process as two-stage tandem queues,
where transmission queue is followed by computation queue.
Note that each device is usually equipped with multiple sensors of various types (e.g., camera, lidar, GPS). 
As for task generation,
multiple sensors with complementary properties are usually fused together to generate a perception task \cite{multi-sensor}. 
Considering the different sampling rates among sensors, 
along with random data fusion delay and the frequency drift of sampling clock crystal oscillator,
we abstract the task generation process as Poisson. 
This assumption facilitates analysis without impacting the essential trade-offs in the investigated problem\cite{poisson}.


As for task offloading, 
at any time,
only one task from the same device can be offloaded.
Similarly,
the offloaded tasks are queued at edge/fog server before they are executed.
We consider a quasi-static scenario where uplink channels and pre-assigned computing power remain unchanged over the time period of observation\cite{channelmodel},
which yields constant service time at both communication and computation queues. 
Note that due to different sampling rates and data sizes among sensors, 
the generated tasks may contain different combinations of multi-modal sampled data,
thus the task size is random. 
Therefore, 
the offloaded tasks experience \emph{i.i.d} random transmission time and computation time. 
Then the transmission queue and computation queue can be modeled as single-server FIFO queue with \emph{i.i.d} service time of different rates.
Both queues are considered to be non-preemptive, 
i.e., a newly arrived task does not replace an older task in the queue.
Moreover,
we also investigate the \emph{zero-wait} task generation policy,
where devices generate tasks only when channel is idle and thus eliminate waiting time in transmission queue.
 
\section{Timeliness of Information Framework}
\label{sec:II:ToIFramework}
In this section, 
we present the mathematical modeling for timeliness of information (ToI), 
which measures the effectiveness of computation offloading by capturing the temporal correlation between status updates extracted from tasks and the real-time status at source devices. 
We find that ToI is determined by queueing delay at both transmission and computation stages. 
Based on the framework presented in this section, 
the closed-form expressions of ToI at edge-tier and fog-tier are derived in following sections (Sec.~\ref{sec:III:edgeTier} and Sec.~\ref{sec:IV:fogTier}), 
where the details of queue dynamics are discussed. 

As for real-time networked systems considered in this paper,
IoT devices generate perception tasks to track dynamics of a physical process (in the form of a stochastic process $X_t$)
and offload them to edge/fog server.
At the server,
status updates are extracted to construct an estimation ($\hat X_t$) of the real-time status ($X_t$).
The \emph{timeliness of information} is defined as the degree of mismatch in relevance to the real-time value of $X_t$.
	
Considering the time-varying nature of monitored physical process,
the timeliness of information is directly related to the time elapsed since the freshest status update is extracted,
which can be captured based on the concept of Age of Information.
In the sequel,
we characterize timeliness of information in two forms:
(1) generalized measures ($\bar \Delta$), where no prior knowledge of the underlying process is available;
and (2) process-related measures ($\bar V$), where the structure of underlying process is leveraged to obtain a more specific indicator.

\subsubsection{Timeliness of Information}
As for real-time causal reconstruction of a general physical process,
a fresher status update tends to yield a better performance.
The real-time reconstruction error is considered to grow linearly with time. 
Therefore, 
we employ the averaged age ($\bar{\Delta}$) as a measure for timeliness of information, 
and extend the conventional age metric to include both transmission and computation stages.
At any time instance $t$,
if a server's most recently executed task was generated at time $u_t$,
the age of obtained status updates is characterized by the random process $\Delta(t)=t-u_t$.
	
Denote $X_n$ as generation interval between tasks,
$\lambda$ as task generation rate,
and $T_n$ as system time of task $n$ (including both transmission and computation).
The timeliness of information is defined as the integral of age function normalized by the time interval of observation ($\mathcal T$) as in \cite{yates-aoi-jsac}:
\begin{equation}\label{eq:age}
		\begin{split}
			\bar{\Delta} &= \lim_{\mathcal{T} \to \infty}\frac{1}{\mathcal{T}}\int_0^{\mathcal{T}}\Delta(t)dt
			= \lambda \left( \mathbb{E}[X_nT_n]+\frac{1}{2}\mathbb{E}[X_n^2]\right).
		\end{split}
\end{equation}

\subsubsection{Process-related Timeliness of Information}

In case when prior knowledge about the structure of monitored process is available,
the correlation properties could be exploited to derive a more meaningful measure for timeliness of information.
To take an example, we focus on stationary Gauss-Markov process $\{X_t\}$ with exponential covariance function {\footnote{Note that the methodology presented in this paper can be applied to various stationary Gaussian processes with integrable covariance functions\cite{kernels}. } and variance as $\mathbb{D}[X_t]=\sigma^2$,
which is widely employed to describe physical activity patterns for networked control tasks
(e.g., environmental sensor measurements, node mobility in ad hoc networks)\cite{ornee-ou}.
Assume that a device samples the process and generates a perception task at time $u_t$,
the offloaded task is executed at time $t$ to extract status update $X_{u_t}$,
which is used to construct an estimate $\hat X_t$ using linear minimum mean square error estimator \cite{lmmse-predict}.
We employ the time-average mean-squared error as an indicator for the timeliness of information:
\begin{equation}\label{eq:value}
	\begin{split}
		\bar{V} &= \lim_{\mathcal{T} \to \infty}\frac{1}{\mathcal{T}} \mathbb{E}\big[\int_0^{\mathcal T}(X_t-\hat{X}_t)^2dt \big]\\
		&=\lim_{\mathcal{T} \to \infty}\frac{1}{\mathcal{T}}\int_0^{\mathcal{T}} \big[\sigma^2-\sigma^2e^{-\kappa|t-u_t|}dt\big]\\
		&=\sigma^2 - \frac{\sigma^2\lambda}{\kappa} (\mathbb{E}[e^{-\kappa T_n}]-\mathbb{E}[e^{-\kappa(X_n+T_n)}]).
	\end{split}
\end{equation}
where $\kappa>0$ is a parameter.
The process-related timeliness of information is characterized as a non-linear function of age.

As shown in expressions (\ref{eq:age}) and (\ref{eq:value}),
the timeliness of information is determined by the sojourn time in both transmission and computation queues. 
In the following Sec. \ref{sec:III:edgeTier} and Sec. \ref{sec:IV:fogTier},
the queue dynamics at edge tier and fog tier will be discussed in detail, respectively. 
	
\section{Timeliness of Information at Edge Tier \\
		M/M/1 -- M/M/1 Tandem}
\label{sec:III:edgeTier}
	
In this section,
we derive closed-form ToI expressions for computation offloading at edge tier.
To realize responsive data processing,
we assume parallel task execution mode, 
where the computation resource at server is partitioned and dedicated to IoT devices using virtualization technique.
Suppose that tasks are generated at each device by a Poisson process with rate $\lambda$,
and the service times of transmission and computation are exponential random variables with rate $u_t$ and $u_c$, respectively.
Both queues are of infinite size and nonpreemptive.
Therefore,
the computation offloading process at edge tier can be abstracted as a M/M/1-M/M/1 tandem queue,
as shown in Fig.~\ref{edge}.
In the following, 
we derive closed-form expressions for ToI in both generalized form and process-related form,
regarding whether the structure of underlying process is available.

\begin{figure}[h]
	\raggedright{\includegraphics[scale = 0.375]{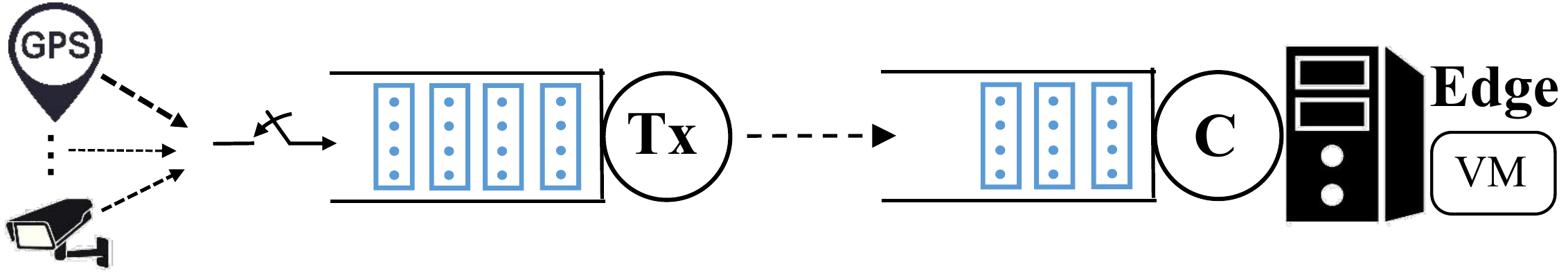}}
	\caption{Queuing model at edge tier.}
	\label{edge}
\end{figure}
	
\subsection{Timeliness of Information Analysis}
\label{sec:subsec:edge-agnostic}

Based on equation \eqref{eq:age},
ToI is characterized as
$\bar{\Delta}= \lambda( \mathbb{E}[X_nT_n]+\frac{1}{2}\mathbb{E}[X_n^2])$.
$X_n$ denotes the task generation interval
with $\mathbb{E}[X_n^2] = 2/\lambda^2$.
Since system time $T_n$  tends to be negatively correlated with $X_n$ (i.e., a smaller task load reduces network congestion),
it complicates the evaluation of ${E}[X_nT_n]$.
As for a general physical process,
the timeliness of information is presented in
the following proposition.
	
\begin{proposition}
Assume M/M/1 for transmission queue followed by M/M/1 for computation queue, 
the timeliness of information is expressed as:
\begin{equation}\label{eq:single average age}
	\begin{split}
				\bar{\Delta}=&\frac{{{\lambda ^2}}}{{\mu _t^2({\mu _t} - \lambda )}} + \frac{1}{{{\mu _t}}} + \frac{{{\lambda ^2}}}{{\mu _c^2({\mu _c} - \lambda )}} \\
				& +\frac{1}{{{\mu _c}}} + \frac{{{\lambda ^2}}}{{{\mu _t}{\mu _c}({\mu _c} + {\mu _c} - \lambda )}} + \frac{1}{\lambda }.
			\end{split}
	\end{equation}
		
		\begin{proof}
			$T_{n,c}$ can be decomposed into four parts,
			including waiting time in transmission queue $W_{n,t}$,
			service time in transmission queue $S_{n,t}$,
			waiting time in computation queue $W_{n,c}$,
			and service time in computation queue $S_{n,c}$:
			\begin{equation}\label{eq:T2}
				T_n=W_{n,t}+S_{n,t}+W_{n,c}+S_{n,c}.
			\end{equation}
			
			$S_{n,t}$ and $S_{n,c}$ are independent from $X_n$, then we have:
			\begin{equation}
				\begin{split} \label{eq:average_AoI}
					\bar{\Delta} =&\lambda \left( \mathbb{E}[X_nT_n]+\frac{1}{2}\mathbb{E}[X_n^2] \right)\\
					=&\lambda \bigg( \mathbb{E}[X_nW_{n,t}]+\mathbb{E}[X_n]\mathbb{E}[S_{n,t}]+\mathbb{E}[X_nW_{n,c}] \\
					&+\mathbb{E}[X_n]\mathbb{E}[S_{n,c}]
					+\frac{1}{2}\mathbb{E}[X_n^2] \bigg)\\
					=&\lambda \bigg[ \mathbb{E}[X_nW_{n,t}]\hspace{-0.3em}+\hspace{-0.3em}\mathbb{E}[X_nW_{n,c}]
					\hspace{-0.3em}+\hspace{-0.3em}\frac{1}{\lambda}\left( \frac{1}{\mu_t}\hspace{-0.3em}+\hspace{-0.3em}\frac{1}{\mu_c}\hspace{-0.3em}+\hspace{-0.3em}\frac{1}{\lambda} \right) \bigg].
				\end{split}
			\end{equation}
			
			Next,
			we detail the analysis of $\mathbb{E}[X_nW_{n,t}]$ and $\mathbb{E}[X_nW_{n,c}]$.
			\\
			
			\subsubsection{Calculation of  {$\mathbb{E}[X_nW_{n,t}]$}}
			\
			
			According to Bayes formula, we have:
			\begin{equation}\label{eq:XW}
				\begin{split}
					\mathbb{E}[X_nW_{n,t}]&=\int_0^{\infty}x\mathbb{E}[W_{n,t}|X_n=x]f_{X_n}(x)dx   \\
					&=\int_0^{\infty} \!\! x\left(\! \int_0^{\infty}\!\! tf_{W_{n,t}|X_n=x}(t)dt \!\right)\!f_{X_n}(x)dx.
				\end{split}	
			\end{equation}
			
			\begin{lemma}
				In M/M/1 for transmission queue, the PDF of $W_{n,t}$ conditioned on $X_n$ is given by:
				\begin{equation}\label{eq:fW|x}
					f_{W_{n,t}|X_n=x}(t)=(\mu_t-\lambda)e^{-(\mu_t-\lambda)(t+x)}~~~~(t>0).
				\end{equation}
				\begin{proof}
					If transmission queue is empty upon generation of task $n$, then $W_{n,t}=0$.
					Otherwise,
					$W_{n,t}$ is the elapsed time from generation of task $n$ until the departure of $(n-1)$-th task.
					Then we evaluate $W_{n,t}$ as:
					\begin{equation} \label{eq:+}
						W_{n,t}=(T_{n-1,t}-X_{n})^+. \nonumber
					\end{equation}
					
					Since $T_{n-1,t}$ and $X_{n}$ are independent, we have:
					\begin{equation}
						\begin{split}\label{eq:Wt|x}
							f_{W_{n,t}|X_n=x}(t)=f_{T_{n,t}}(t\!+\!x)
							=(\mu_t\!-\!\lambda)e^{-(\mu_t\!-\!\lambda)(t+x)}~(t>0). \nonumber
						\end{split}
					\end{equation}
				\end{proof}
			\end{lemma}
			
			Combining \eqref{eq:XW} and \eqref{eq:fW|x},
			we have:
			\begin{equation}\label{eq:XW_t}
				\begin{split}
					\mathbb{E}[X_nW_{n,t}]=\frac{\lambda}{\mu_t^2(\mu_t-\lambda)}.
				\end{split}
			\end{equation}
			\\
			\subsubsection{Calculation of {$\mathbb{E}[X_nW_{n,c}]$}}
			\
			
			As before, the joint expectation can be obtained as:
			\begin{equation}
				\begin{split}\label{eq:XWc}
					\hspace{-2mm} \mathbb{E}[X_nW_{n,c}]
					=\int_0^{\infty}\!x\left(\!\int_0^{\infty}\!tf_{W_{n,c}|X_n=x}(t)dt\!\right)\!f_{X_n}(x)dx.
				\end{split}
			\end{equation}
			
			Since there is no packet replacement policy adopted between transmission and computation stages,  the arrival of computation queue is statistically identical to the departure of transmission queue.
			We introduce $D_{n,t}$ as intermediate variable,
			which denotes the inter-departure time of transmission queue.
			Then we have:
			\begin{equation}\label{eq:fWc|x}
				\hspace{-0.2mm} {f_{{W_{n,c}}|{X_n} = x}}(t) \!=\! \int_0^\infty \!\! {{f_{{D_{n,t}}|{X_n} = x}}} (y){f_{{W_{n,c}}|{D_{n,t}} = y}}(t)dy.
			\end{equation}
			
			Since $D_{n,t}$ is statistically identical with $X_{n}$,
			following Lemma 1, we have:
			\begin{equation}\label{eq:W-D}
				{f_{{W_{n,c}}|{D_{n,t}} = y}}(t) = (\mu_c-\lambda){e^{ - (\mu_c-\lambda)(t+y)}}~~~(t>0).
			\end{equation}
			
			As for the transmission stage,
			we need to derive the PDF of $D_{n,t}$ conditioned on $X_{n}$.
			We consider two conditions: transmission queue is idle, or busy upon generation of task $n$:
			\begin{equation}\label{eq:fDx_fenjie}
				\begin{split}
					{f_{{D_{n,t}}|{X_n} = x}}(t)
					=&{P_{busy,t}}{f_{{D_{n,t}}|{W_{n,t}} > 0,{X_n} = x}}(t)\\
					&+ {P_{idle,t}}{f_{{D_{n,t}}|{W_{n,t}} = 0,{X_n} = x}}(t).  
				\end{split}
			\end{equation}
			
			\textbf{a) Busy:}
			The probability of transmission queue being busy upon generation of task $n$ can be calculated as:
			\begin{equation}
				\begin{split}\label{eq:Pib}
					P_{busy,t}&=P(W_{n,t}>0|X_n=x)= P(T_{n-1,t}>x)  \\
					&=e^{-(\mu_t-\lambda)x}.
				\end{split}
			\end{equation}
			
			Since $D_{n,t}=S_{n,t}$, we have:
			\begin{equation}\label{}
				f_{D_{n,t}|W_{n,t}>0, X_n=x}(t)=f_{S_{n,t}}(t)=\mu_t e^{-\mu_t t}.  
			\end{equation}
			
			\textbf{b) Idle:} The probability of transmission queue being idle upon generation of task $n$ can be calculated as:
			\begin{equation}
				\begin{split}\label{eq:Pii}
					P_{idle,t}&=1-P_{busy,t}=1-e^{-(\mu_t-\lambda)x}.
				\end{split}
			\end{equation}
			
			\begin{lemma}
				In M/M/1 for transmission queue,
				the PDF of $D_{n,t}$ conditioned on $X_n$ and $W_{n,t}$ is given by:
				\begin{equation}\label{eq:D|WX}
					\begin{split}
						&{f_{{D_{n,t}}|{W_{n,t}}= 0,{X_n} = x}}(t) \\
						= & \left\{ {\begin{array}{*{20}{l}}
								{\frac{1}{{{P_{idle,t}}}}\frac{{\mu_t-\lambda}}{{2 - {\rho _t}}}{e^{ - (\mu_t-\lambda)x}}\left[{e^{(\mu_t-\lambda)t}} - {e^{ - {\mu _t}t}}\right],{\rm{if}}\;{\rm{t}} \le {\rm{x}}}\\
								{\frac{1}{{{P_{idle,t}}}}\frac{{\mu_t-\lambda}}{{2 - {\rho _t}}}\left[{e^{{\mu _t}x}} - {e^{ - (\mu_t-\lambda)x}}\right]{e^{ - {\mu _t}t}},{\rm{if}}\;{\rm{t}} > {\rm{x}}}.
						\end{array}} \right.   
					\end{split}
				\end{equation}
				\begin{proof}
					The detailed proof is given in Appendix A.
				\end{proof}		
			\end{lemma}
		
			Then, combining \eqref{eq:fDx_fenjie}--\eqref{eq:D|WX}, the PDF of $D_{n,t}$ conditioned on $X_n$ can be calculated as
			\begin{equation}
				\begin{split}\label{eq:fDx}
					f&_{{D_{n,t}}|{X_n} = x}(t) =\mu_t e^{-\mu_t t}e^{-(\mu_t-\lambda)x}\\
					&+ \left\{ {\begin{array}{*{20}{l}}
							{\frac{{\mu_t-\lambda}}{{2 - {\rho _t}}}{e^{ - (\mu_t-\lambda)x}}\left[{e^{(\mu_t-\lambda)t}} - {e^{ - {\mu _t}t}}\right],{\rm{if}}\;{\rm{t}} \le {\rm{x}}}\\
							{\frac{{\mu_t-\lambda}}{{2 - {\rho _t}}}\left[{e^{{\mu _t}x}} - {e^{ - (\mu_t-\lambda)x}}\right]{e^{ - {\mu _t}t}},{\rm{if}}\;{\rm{t}} > {\rm{x}}}.
					\end{array}} \right.
				\end{split}
			\end{equation}
			
			Further,  by putting \eqref{eq:fWc|x}, \eqref{eq:W-D} and \eqref{eq:fDx} in \eqref{eq:XWc}, we have:
			\begin{equation}\label{eq:XW_c}
				\mathbb{E}[X_nW_{n,c}]=\frac{\lambda }{{\mu _c^2({\mu _c} - \lambda )}} + \frac{\lambda }{{{\mu _t}{\mu _c}({\mu _t} + {\mu _c} - \lambda )}}.
			\end{equation}
			
			Finally,
			by putting \eqref{eq:XW_t} and \eqref{eq:XW_c} in \eqref{eq:average_AoI},
			we can draw the conclusion in Proposition 1, and get the closed-form expression for timeliness of information at edge tier.			
		\end{proof}		
	\end{proposition}

	\noindent\textbf{Zero-wait Task Generation:}
	We investigate \emph{zero-wait} task generate policy as a special case,
	where an acknowledgement (ACK) is fed back to the device once a task is offloaded.
	Then the device has access to the idle/busy state of server in real-time,
	and generates a new task once the channel is idle.
	It would seem to be a better task generation policy,
	as it eliminates the waiting time in transmission queue.
	However,
	it does not always yields the best performance in tandem queue,
	which will be discussed in Sec. \ref{sec:subsec:II-validation}.
	
	\begin{corollary}
		Under zero-wait policy,
		the timeliness of information in M/M/1-M/M/1 is expressed as:
		\begin{equation}\label{}
			\begin{split}
				&\bar{\Delta}^* = \frac{2}{\mu_t}+\frac{1}{\mu_c}+\frac{\mu_t}{\mu_c(\mu_c-\mu_t)}. \nonumber
			\end{split}
		\end{equation}
		
		\begin{proof1}
			Given task generation interval $X_n=S_{n-1,t}$,
			and inter-departure time $D_{n,t}=S_{n,t}$,
			\eqref{eq:average_AoI} can be written as:	
			\begin{equation}
				\begin{split} \label{eq:age_peaK}
					&\bar{\Delta}^* = \mathbb{E}[S_{n,t}]+\mathbb{E}[S_{n,c}]+\frac{1}{2}\frac{\mathbb{E}[S_{n,t}^2]}{\mathbb{E}[S_{n,t}]}+\mathbb{E}[W_{n,c}], \nonumber
				\end{split}
			\end{equation}
			where
			$\mathbb{E}[S_{n,t}]=1/\mu_t$, $\mathbb{E}[S_{n,t}^2]=2/\mu_t^2$, $\mathbb{E}[S_{n,c}]=1/\mu_c$.
			
			following Lemma 1, the conditioned PDF of waiting time in computation queue $\mathbb{E}[W_{n,c}]$ is:
			\begin{equation}\label{eq:W-S}
				\begin{split}
					{f_{{W_{n,c}}|{D_{n,t}} = y}}(t) &= {f_{{W_{n,c}}|{S_{n,t}} = y}}(t)\\ &=(\mu_c-\mu_t){e^{ - (\mu_c-\mu_t)(t+y)}}~~~(t>0).
				\end{split}
			\end{equation}
			
			Then we have:
			\begin{align}\label{eq:Wc}
				\mathbb{E}[W_{n,c}]
				=&\int_0^{\infty} \left( \int_0^{\infty}t{f_{{W_{n,c}}|{S_{n,t}} = y}}(t)dt \right)f_{S_{n,t}}(y)dy  \nonumber \\
				=&\frac{\mu_t}{\mu_c(\mu_c-\mu_t)}.    \nonumber 
				\quad\quad\quad\quad\quad\quad\quad\quad\quad\quad\quad\quad\quad\quad  \hfill \blacksquare 
			\end{align}
		\end{proof1}
	\end{corollary}
	
	\subsection{Process-related Timeliness of Information Analysis}
	
	Based on equation \eqref{eq:value}, 
	the process-related ToI is characterized as 
	$\bar{V}=\sigma^2 - \frac{\sigma^2\lambda}{\kappa} (\mathbb{E}[e^{-\kappa T_n}]-\mathbb{E}[e^{-\kappa(X_n+T_n)}])$.
	The nonlinearity induces inseparable terms consisting of stochastic waiting times at both communication and computation stages,
	which complicates the analysis.
	The process-related timeliness of information is presented in
	the following proposition.
			
\begin{proposition}
	In M/M/1 for transmission queue, followed by M/M/1 for computation queue,
	the process-related timeliness of information is:
	\begin{equation}\label{}
		\begin{split}
			\bar V\!=&\sigma^2 \!-\!  \frac{{\sigma^2 \lambda ({\mu _t} - \lambda )({\mu _c} - \lambda )}}{{\kappa(\kappa + {\mu _t} - \lambda )(\kappa + {\mu _c} - \lambda )}}
			\!+\!\frac{{{\sigma^2 \lambda^2\mu _t\mu_c}}}{({\kappa + {\mu _t}})(\kappa+\mu_c)}\\
			&\times \left[\frac{1}{\kappa({\kappa + \lambda })}- \frac{{ 1}}{{(\kappa + {\mu _t} - \lambda )(\kappa  + {\mu _c} - \lambda )}}\right. \\
			&\left. + \frac{\lambda}{{(\kappa + {\mu _c} - \lambda )(\kappa + {\mu _t} - \lambda )(\kappa + {\mu _t} + {\mu _c} - \lambda )}} \right. \\
			&\left. - \frac{{ ({\mu _c} - \lambda )}}{{(\kappa + {\mu _t} - \lambda )(\kappa + {\mu _t})(\kappa + {\mu _t} + {\mu _c} - \lambda )}} \right. \\
			&\left. - \frac{{({\mu _t} - \lambda )}}{{(\kappa + {\mu _c} - \lambda )(\kappa + {\mu _c})(\kappa + {\mu _t} + {\mu _c} - \lambda )}} \right] . \nonumber
		\end{split}
	\end{equation}
		
		\begin{proof}
			As defined in \eqref{eq:value},
			$\bar{V}=\sigma^2 - \frac{\sigma^2\lambda}{\kappa} (\mathbb{E}[e^{-\kappa T_n}]-\mathbb{E}[e^{-\kappa(X_n+T_n)}])$.
			In the following,
			we detail the analysis of $\mathbb{E}[e^{-\kappa T_n}]$ and $\mathbb{E}[e^{-\kappa(X_n+T_n)}]$.
			\\
			\subsubsection{Calculation of  {$\mathbb{E}[e^{-\kappa T_n}]$}}
			\
			
			Since $T_{n,t}$ and $T_{n,c}$ are independent, the PDF of $T_n$ is:
			\begin{equation}\label{eq:V_T}
				\begin{split}
					\mathbb{E}[e^{-\kappa T_n}]=&\int_0^\infty  {{e^{ - \kappa t}}} \left[{f_{{T_{n,t}}}}(t) \otimes {f_{{T_{n,c}}}}(t)  \right]dt \\
					=&\frac{{\left( {{\mu _t} - \lambda } \right)\left( {{\mu _c} - \lambda } \right)}}{{\left( {\kappa + {\mu _t} - \lambda } \right)\left( {\kappa + {\mu _c} - \lambda } \right)}}.
				\end{split}
			\end{equation}
			\\
			
			\subsubsection{Calculation of  {$\mathbb{E}[e^{-\kappa(X_n+T_n)}]$}}
			\
			
			As specified in \eqref{eq:T2},
			$T_{n}$ consists of $W_{n,t}$ , $S_{n,t}$, $W_{n,c}$ and $S_{n,c}$,
			among which only  $S_{n,c}$ is independent to the other three variables and $X_n$.
			Then we have:
			\begin{equation}\label{eq:nadir V}
				\mathbb{E}[e^{-\kappa(X_n\!+\!T_n)}]\!=\!\mathbb{E}[e^{-\kappa(X_n+\!W_{n,t}+\!W_{n,c}+\!S_{n,t})}] \mathbb{E}[e^{\kappa S_{n,c}}],
			\end{equation}
			where the second term can be derived using the PDF of $S_{n,c}$.
			As for the first term, it can be written as: 
			\begin{equation}
				\begin{split}\label{eq:V_other1}
					&\mathbb{E}[e^{-\kappa(X_n+\!W_{n,t}+\!W_{n,c}+\!S_{n,t})}]\\
					\hspace{-5mm} =\!&\int_0^\infty \int_0^\infty \!\! {e^{ - \kappa s}}e^{ - \kappa x}\mathbb{E}[{e^{ - \kappa (W_{\!n,t}+W_{\!n,c})}}|S_{n,t}\!\!=\!s,\! X_{\!n}\!\!=\!x] \\
					& \quad\quad\quad\quad  {f_{X_n}}\!(x){f_{S_{n,t}}}\!(s)dxds,
				\end{split}
			\end{equation}
			in which
			\begin{equation}
				\begin{split}\label{eq:V_other3}
					&\mathbb{E}[{e^{ - \kappa(W_{n,t}+W_{n,c})}}|S_{n,t}=s,X_n=x]\\
					=&\int_0^\infty  {e^{ - \kappa t}} {f_{{W_{n,t}} + {W_{n,c}}|{X_n} = x,{S_{n,t}} = s}}(t)dt.
				\end{split}
			\end{equation}
			
			We consider four possible combinations of $W_{n,t}$ and $W_{n,c}$ as shown in Fig.~\ref{4 cases}.
			We introduce the inter-departure time $D_{n,t}$ of transmission queue as an intermediate variable to characterize the dependency of variables in both queues. 
			
			\begin{figure}[h]
				\centerline{\includegraphics[scale = 0.45]{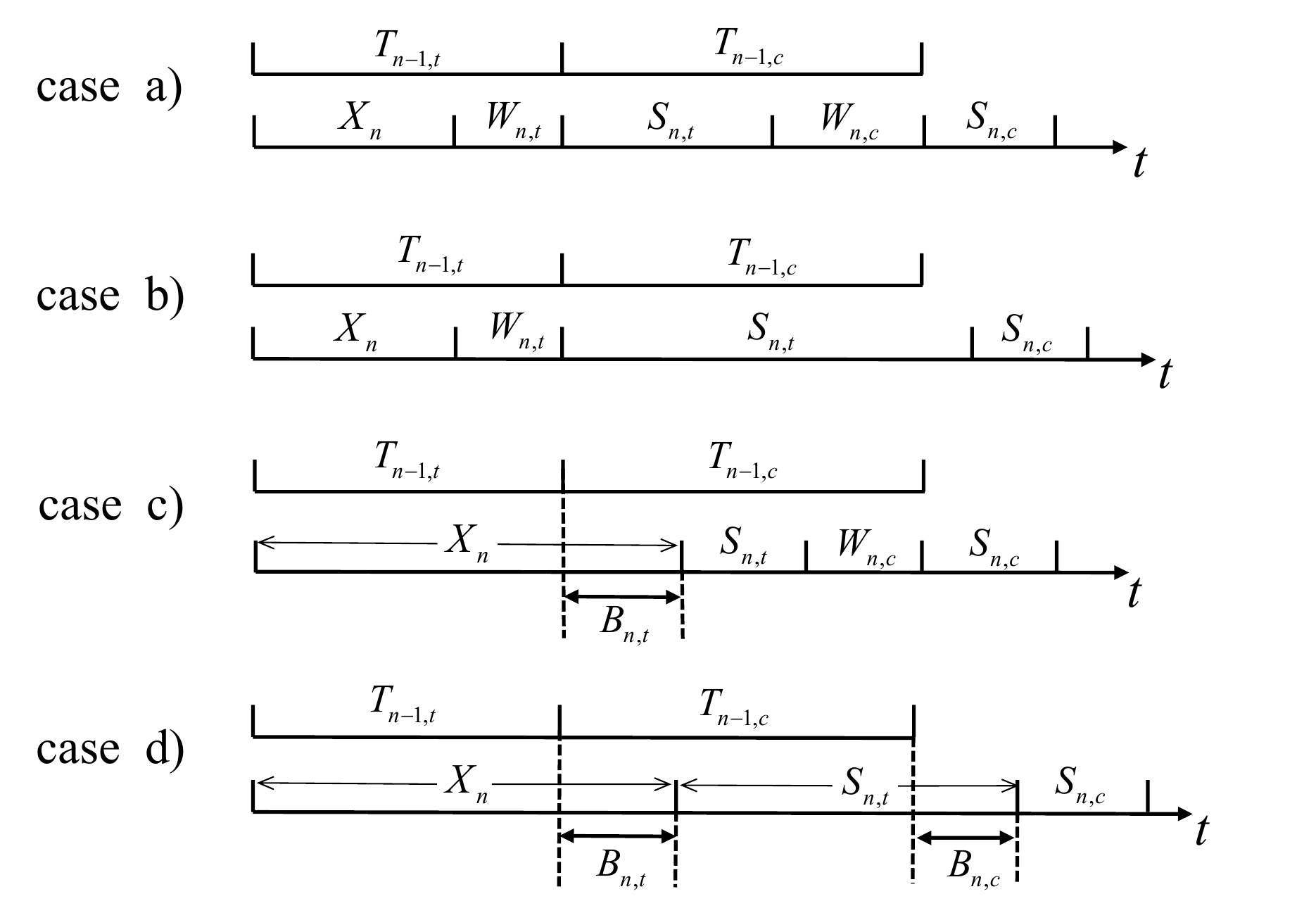}}
				\caption{Schematic of possible combinations of $W_{n,t}$ and $W_{n,c}$. Case a): task $n$ waits in both queues. Case b): task $n$ waits only in the first queue. Case c): task $n$  waits only in the second queue. Case d): task $n$ is immediately served in both queue.}
				\label{4 cases}
			\end{figure}
			
			\textbf{Case a):\ }
			When both transmission and computation queue are busy, $W_{n,t}>0, W_{n,c}>0$,
			the inter-departure time $D_{n,t}$ is statistically identical to the service time $S_{n,t}$,
			while $W_{n,c}$ and $W_{n,t}$ are independent.
			We evaluate the conditioned PDF of $(W_{n,t}+W_{n,c})$ as
			the convolution of conditioned PDF of $W_{n,t}$ and $W_{n,c}$.
			Then based on \eqref{eq:fW|x} and \eqref{eq:W-D}, we have:
			\begin{equation}
				\begin{split}\label{eq:Wc>0}
					&{f_{{W_{n,t}} + {W_{n,c}}|{W_{n,c}} > 0,{W_{n,t}} > 0,{X_n} = x,{S_{n,t}} = s}}(t)\\
					=&{f_{{W_{n,t}}|{W_{n,t}} > 0,{X_n} = x}}(t) \otimes {f_{{W_{n,c}}|W_{n,c}>0, {D_{n,t}} = s}}(t)\\		
					= &\frac{{({\mu _t} - \lambda )({\mu _c} - \lambda )}}{{{\mu _t} - {\mu _c}}} \left[ - {e^{ - (\mu_t-\lambda)t}} + {e^{ - (\mu_c-\lambda)t}}\right].
				\end{split}
			\end{equation}
			
			\textbf{Case b):\ }
			When communication queue is busy and computation queue is idle ($W_{n,t}>0,W_{n,c}=0$),
			$W_{n,t}$ and $S_{n,t}$ are independent.
			Based on \eqref{eq:fW|x} and \eqref{eq:Pib},
			the conditioned PDF of $(W_{n,t}+W_{n,c})$ can be derived as:
			\begin{equation}
				\begin{split} \label{eq:Wc=0}
					&f_{{W_{n,t}+W_{n,c}}|W_{n,c}=0,{W_{n,t}} > 0,{X_n} = x,{S_{n,t}}=s}(t)  \\
					& \quad\quad\quad\quad\quad\quad = \frac{f_{{W_{n,t}}|{X_n} = x}(t)}{{{P_{busy,t}}}} \\
					& \quad\quad\quad\quad\quad\quad = (\mu_t-\lambda){e^{ - (\mu_t-\lambda)t}}.
				\end{split}
			\end{equation}
			
			\textbf{Case c):\ }
			When communication queue is idle and computation queue is busy
			($W_{n,t}=0,W_{n,c}>0$),
			we use inter-departure time $D_{n,t}$ to evaluate the dependency among $W_{n,c}$ and $S_{n,t}$, $X_{n}$.
			
			\begin{lemma}
				Given $W_{n,t}=0,W_{n,c}>0$, the PDF of $(W_{n,t}+W_{n,c})$ conditioned on $X_n$ and $S_{n,t}$ is:
				\begin{align} \label{eq:Wt=0,Wc>0}
					&{f_{{W_{n,t}+W_{n,c}}|W_{n,c}>0,{W_{n,t}} = 0,{X_n} = x,{S_{n,t}} = s}}(t)  \\
					\hspace{-5mm} = &\frac{(\mu_c \!\!-\!\lambda)\!(\mu_t \!-\!\lambda){e^{ - (\mu_c\!-\!\lambda)s}}}{P_{idle,t}}\frac{e^{ - (\mu_c\!-\!\lambda)x}\!-\!e^{ - (\mu_t\!-\!\lambda)x}}{\mu_t \!-\! \mu_c}{e^{ - (\mu_c\!-\!\lambda)t}}  \nonumber.
				\end{align}
				
				\begin{proof}
						The proof is given in Appendix B. 
				\end{proof}			
			\end{lemma}
			
			Based on the above three cases,
			we can obtain the conditioned PDF of positive valued $(W_{n,t}+W_{n,c})$.
			
			\begin{lemma}
				Given $W_{n,t}+W_{n,c}>0$, the PDF of $(W_{n,t}+W_{n,c})$ conditioned on $X_n$ and $S_{n,t}$ is:
				\begin{equation}\label{eq:fW}
					\begin{split}
						&{f_{{W_{n,t}} + {W_{n,c}}|{X_n} = x,{S_{n,t}} = s,W_{n,t}+W_{n,c}>0}}(t)\\
						=& \frac{{{\mu _t} - \lambda }}{{{\mu _t} - {\mu _c}}}{e^{ - (\mu_c-\lambda)s}}[({\mu _c} - \lambda ){e^{ - (\mu_c-\lambda)(x+t)}}\\
						&- ({\mu _t} - \lambda ){e^{ - (\mu_t-\lambda)(x+t)}}] + ({\mu _t} - \lambda ){e^{ - (\mu_t-\lambda)(x+t)}}.
					\end{split}
				\end{equation}
				
				\begin{proof}
					The proof is given in Appendix C.
				\end{proof}
			\end{lemma}
			
			\textbf{Case d):\ }
			Then we evaluate the special case when $W_{n,t}+W_{n,c}=0$.
			Based on Lemma 4 and the total probability theorem,
			we have:
			\begin{align}\label{eq:W=0}
				&P(W_{n,t}+W_{n,c}=0|{X_n} = x,{S_{n,t}} = s) \nonumber \\
				=& 1 \!-\! \int_0^\infty  {f_{{W_{n,t}} + {W_{n,c}}|{X_n} = x,{S_{n,t}} = s,W_{n,t}+W_{n,c}>0}}(t) dt \\
				=&\frac{{{\mu _t} \!-\! \lambda }}{{{\mu _t} \!-\! {\mu _c}}}{e^{ -\! (\mu_c-\lambda)s}}[{e^{ -\!(\mu_t-\lambda)x}} \!-\! {e^{ -\!(\mu_c-\lambda)x}}] \!-\! {e^{ -\!(\mu_t-\lambda)x}} \!+\! 1. \nonumber
			\end{align}
			
			According to the analysis of four possible combinations,
			we evaluate \eqref{eq:V_other3} based on \eqref{eq:fW} and \eqref{eq:W=0}:
			\begin{align} \label{eq:V_other4}
				&\mathbb{E}[{e^{ - \kappa (W_{n,t}+W_{n,c})}}|S_{n,t}=s,X_n=x]  \nonumber \\
				=&P(W_{n,t}+W_{n,c}=0|{X_n} = x,{S_{n,t}} = s) \\
				&+ \int_0^\infty  {e^{ - \kappa t}} {f_{{W_{n,t}} + {W_{n,c}}|{X_n} = x,{S_{n,t}} = s,W_{n,t}+W_{n,c}>0}}(t)dt.   \nonumber
			\end{align}

			Then $\mathbb{E}[e^{-\kappa(X_n+T_n)}]$ can be evaluated by combining \eqref{eq:nadir V}, \eqref{eq:V_other1}, \eqref{eq:V_other4} and the PDF of $S_{n,c}$.

			Finally, putting the derived result of $\mathbb{E}[e^{-\kappa(X_n+T_n)}]$ and \eqref{eq:V_T} in \eqref{eq:value}, we can draw the conclusion in Proposition 2,
			and get the final expression for process-related timeliness of information.				
		\end{proof}
	\end{proposition}

	\noindent\textbf{Zero-wait Task Generation:}
	We also evaluate process-related timeliness of information under zero-wait task generate policy.
	
	\begin{corollary}
		Under zero-wait policy,
         the process-related timeliness of information in M/M/1-M/M/1 is expressed as:
		\begin{equation}\label{}
	\begin{split}
		\bar{V}^* =&\sigma^2 - \frac{{{\sigma^2 \mu _t^2  } \left( {{\mu _c} - \mu_t } \right)}}{\kappa  {\left( {\kappa + {\mu _t}  } \right)\left( {\kappa + {\mu _c} - \mu_t } \right)}}\\
		&+\frac{\sigma^2 \mu_t^3\mu_c(\mu_c-\mu_t)}{\kappa  (\kappa+\mu_t)(\kappa+\mu_c)^2(\kappa+\mu_c-\mu_t)}. \nonumber
	\end{split}
\end{equation}
		
		\begin{proof}
			Given task generation interval $X_n=S_{n-1,t}$,
			and inter-departure time $D_{n,t}=S_{n,t}$,
			\eqref{eq:value} can be written as:	
		\begin{equation}\label{eq:V*}
		\begin{split}
			\bar{V}^*
			=&\sigma^2 - \frac{\sigma^2 \mu_t}{\kappa } \Big(\mathbb{E}[e^{-\kappa T_n}]\!-\!\mathbb{E}[e^{-\kappa (W_{\!n,c}+S_{n,t})}]  \\
			& \times \mathbb{E}[e^{\kappa S_{n-1,t}}] \mathbb{E}[e^{\kappa S_{n,c}}] \Big).
		\end{split}
	\end{equation}
			where $ \mathbb{E}[e^{\kappa S_{n-1,t}}]$ and $ \mathbb{E}[e^{\kappa S_{n,c}}]$ can be obtained by PDFs of $S_{n,t}$ and $S_{n,c}$.
			
			As for $\mathbb{E}[e^{-\kappa T_n}]$,
			since $S_{n,t}$ and $T_{n,c}$ are independent, it can be evaluated as:
			\begin{equation}\label{eq:kT}
				\begin{split}
					\mathbb{E}[e^{-\kappa T_n}]=&\int_0^\infty  {{e^{ - \kappa t}}} \left[ {f_{{S_{n,t}}}}(t) \! \otimes \! {f_{{T_{n,c}}}}(t) \right] dt \\
					=&\frac{{{{\mu _t}  } \left( {{\mu _c} - \mu_t } \right)}}{{\left( {\kappa + {\mu _t}  } \right)\left( {\kappa + {\mu _c} - \mu_t } \right)}}.
				\end{split}
			\end{equation}
			
			Then we evaluate $\mathbb{E}[e^{-\kappa (W_{\!n,c}+S_{n,t})}]$ as:
			\begin{align}\label{eq:kWS}
				&\mathbb{E}[{e^{ - \kappa(W_{n,c}+S_{n,t})}}] \nonumber \\
				&\quad\quad= \int_0^\infty  {{e^{ - \kappa y}}} \left( \int_0^\infty  {{e^{ - \kappa t}}} {f_{{W_{n,c}}|{S_{n,t}} = y}}(t)dt \right){f_{{S_{n,t}}}}(y)dy \nonumber \\
				&\quad\quad= \frac{\mu_t(\mu_c-\mu_t)}{(\kappa+\mu_c)(\kappa+\mu_c-\mu_t)}.
			\end{align}
			
			Finally, by putting \eqref{eq:kT} \eqref{eq:kWS} into \eqref{eq:V*}, the closed-form expression of $\bar{V}^*$ is obtained.		
		\end{proof}
	\end{corollary}

	\subsection{Numerical Results and Analysis}
	\label{sec:subsec:II-validation}
	In this section,
	we present numerical results to verify the accuracy of theoretically derived ToI expressions. 
	Moreover, 
    we demonstrate the impact of various system parameters,
	including task generation rate, transmission rate and computation rate.
	We also present the achievable performance under zero-wait task generation policy.
	
\begin{figure*}
	\begin{minipage}{0.33\textwidth}
		\raggedright
		\includegraphics[width=\textwidth]{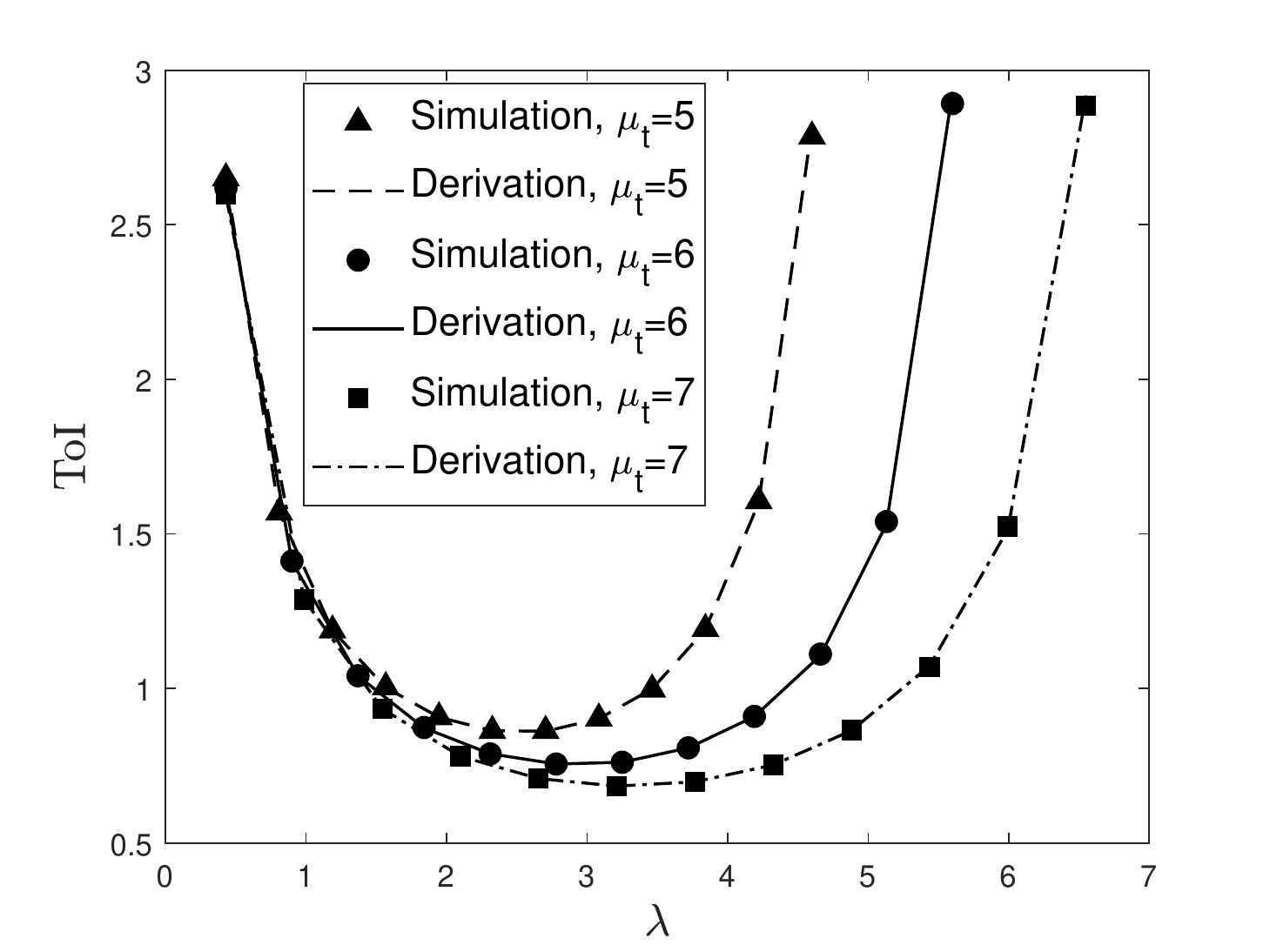}
		\caption{ToI as task generation rate increases under different transmission rates. $\mu_c=8$.}
		\label{AoI_monte}
	\end{minipage}
	\hfill
	\begin{minipage}{0.33\textwidth}
		\centering
		\includegraphics[width=\textwidth]{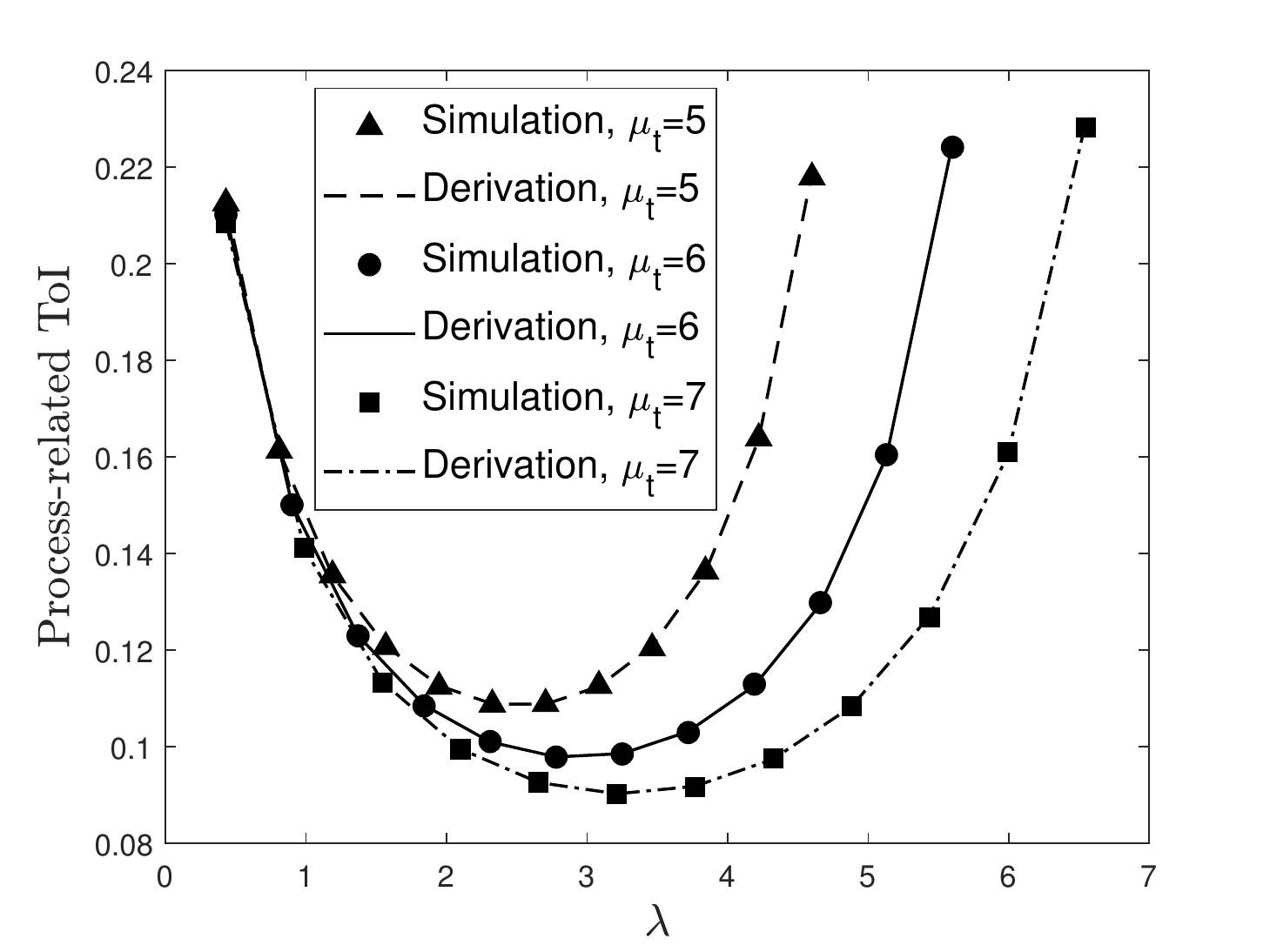}
		\caption{Process-related ToI as task generation rate increases under different transmission rates. $\mu_c=8$.}
		\label{voi_monte}
	\end{minipage}
	\hfill
	\begin{minipage}{0.313\textwidth}
		\raggedleft
		\includegraphics[width=\textwidth]{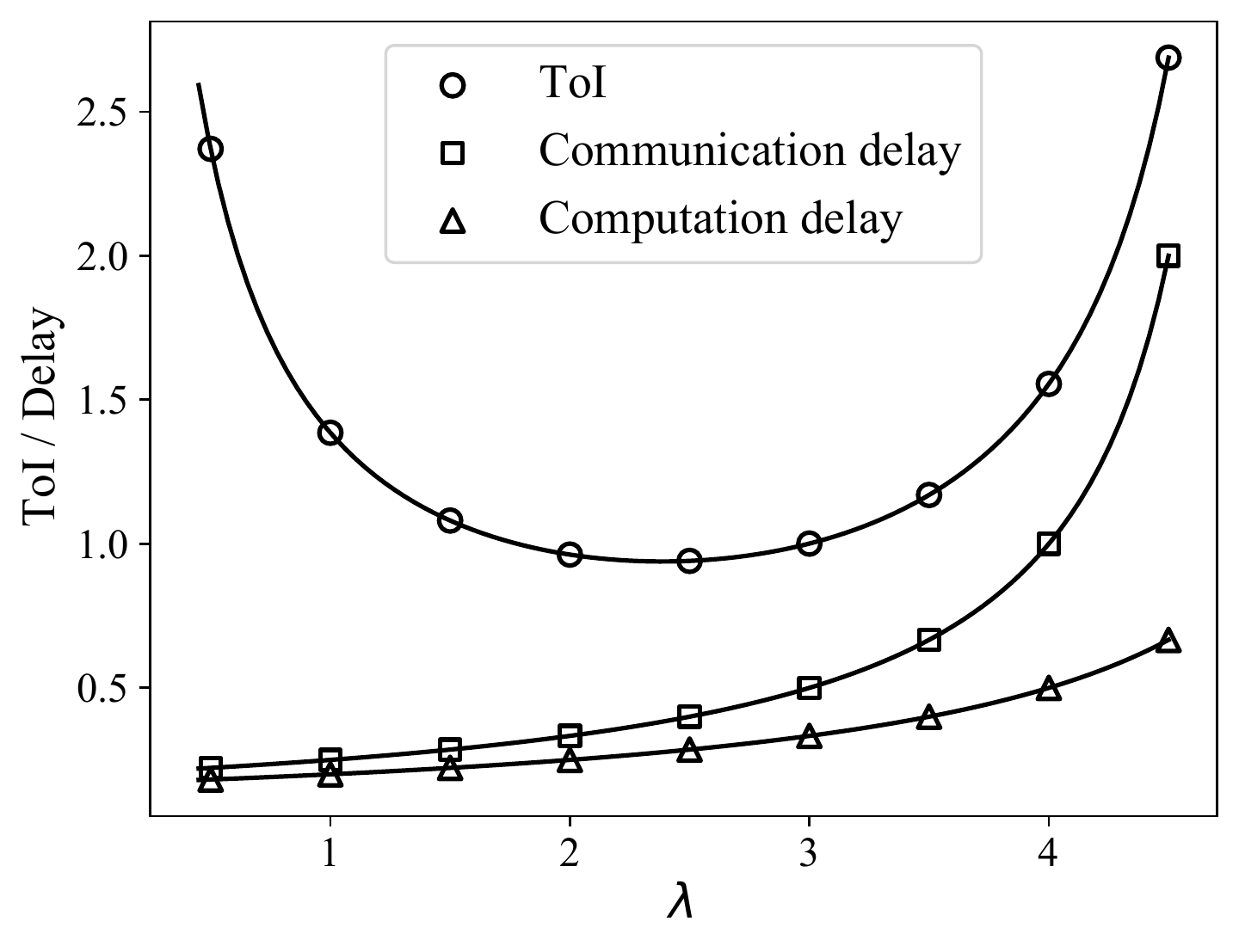}
		\caption{ToI, transmission delay  and computation delay as task generation rate increases. $\mu_t \!=5, \mu_c \!=6$.}
		\label{AoI_delay}
	\end{minipage}
	\begin{minipage}{0.306\textwidth}
		\vspace{3pt}
		\raggedright
		\includegraphics[width=\textwidth]{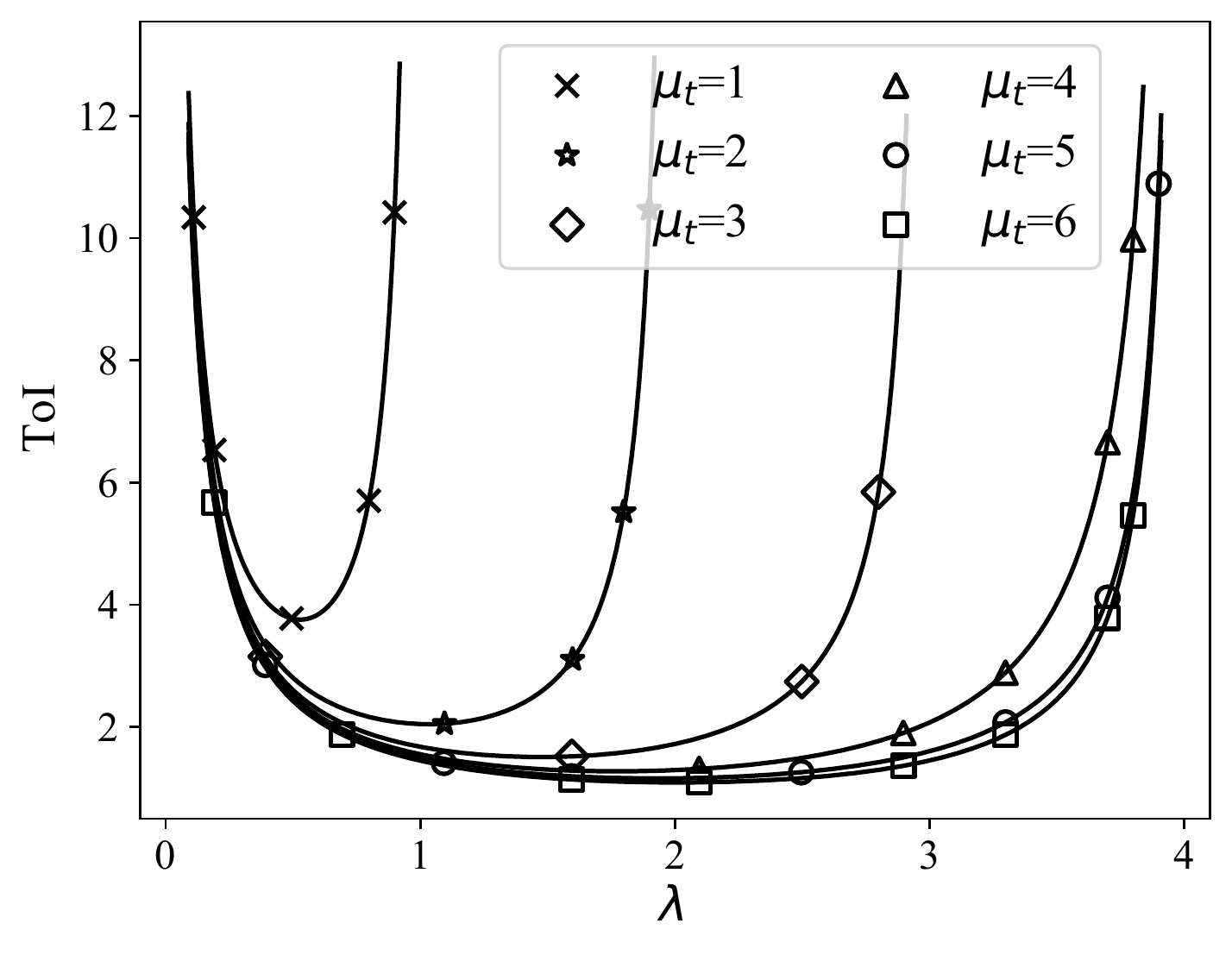}
		\caption{ToI under different transmission rates as task generation rate increases. $\mu_c=4$.}
		\label{AoI_different_utuc}
	\end{minipage}
	\hfill
	\begin{minipage}{0.305\textwidth}
		\vspace{3pt}
		\centering
		\includegraphics[width=\textwidth]{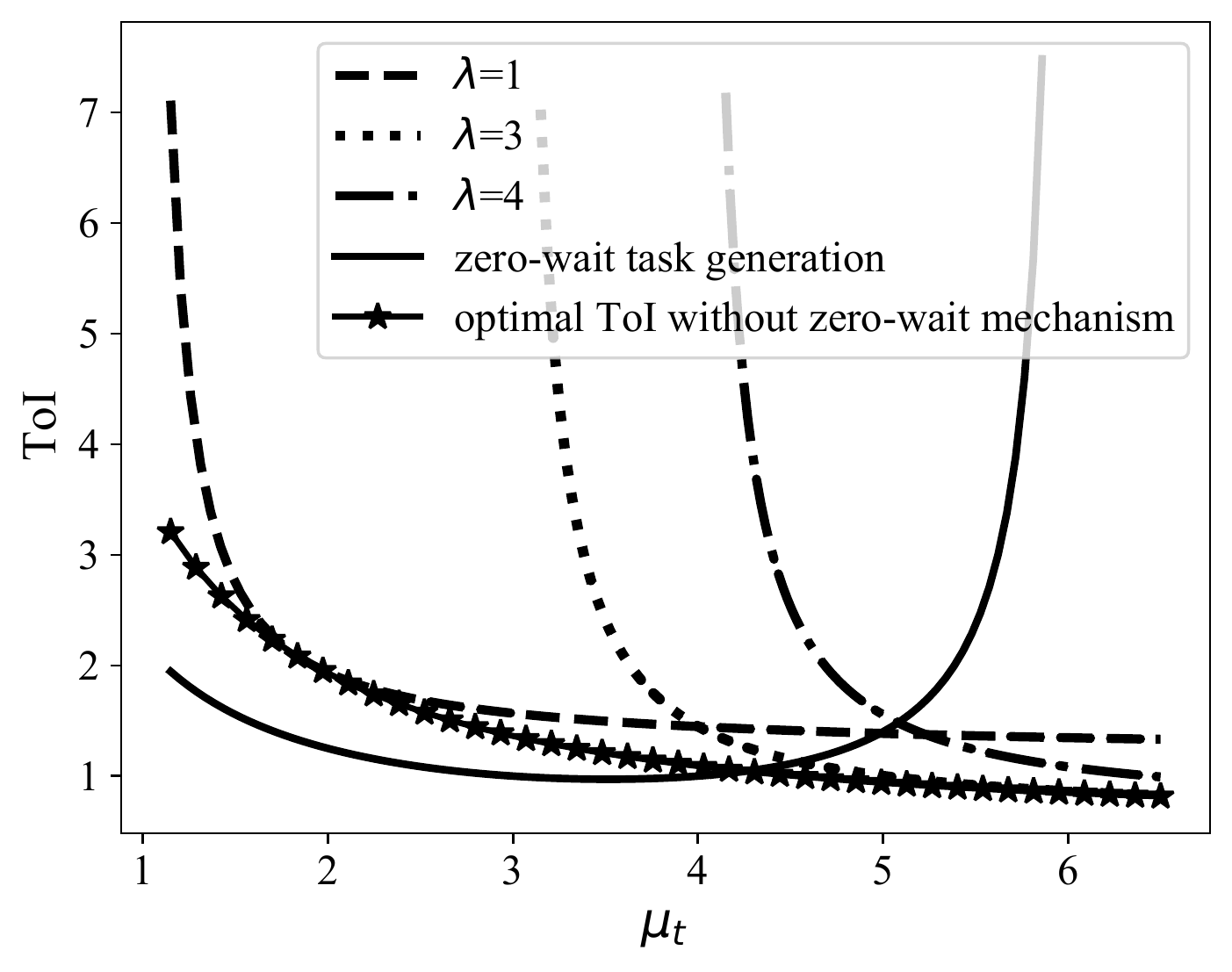}
		\caption{ToI under different task generation policies as transmission rate increases. $\mu_c=6$.}
		\label{AoI_awareness}
	\end{minipage}
	\hfill
	\begin{minipage}{0.31\textwidth}
		\vspace{3pt}
		\raggedleft
		\includegraphics[width=\textwidth]{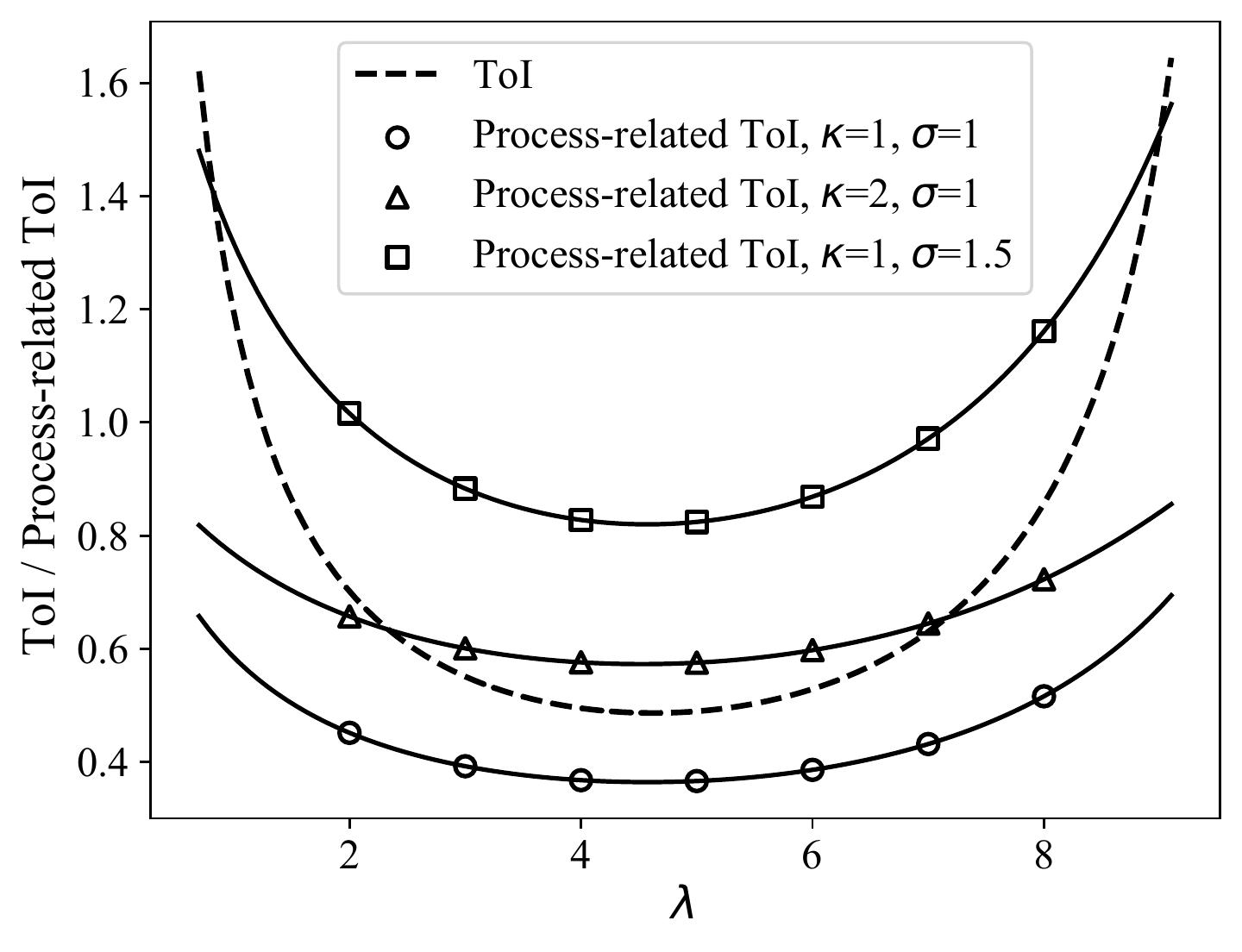}
		\caption{Process-related ToI VS ToI as task generation rate increases, under various parameters of monitored process.}
		\label{compare}
	\end{minipage}
\end{figure*}
	
First, we compare the results of our analysis with Monte Carlo simulations, generating $N=10^6$ tasks and computing ToI and process-related ToI for each. 
The initial stages of each simulation are discarded to ensure that the system have reached a steady state. 
The Monte Carlo simulation consists of a single episode, with all tasks being offloaded one after the other.
As shown in Fig.~\ref{AoI_monte} and Fig.~\ref{voi_monte}, 
since no approximation is involved,
the simulation results perfectly match the theoretically derived curves for both ToI and process-related ToI, 
verifying the soundness of our derivation.
	
Then, we present the difference between timeliness of information with traditional delay metric.
	Fig.~\ref{AoI_delay} shows the trend of ToI,
	transmission delay and computation delay as the task generation rate increases from 0.5 to 4.5,
	under fixed transmission and computation rates.
	As shown in the figure,
	both transmission and computation delay increases monotonically as the increase of task load.
	However,
    ToI first decreases and then increases.
	Note that in real-time networked control systems,
	a sporadic task generation at devices leads to out-of-date perceptions at edge server,
	which is deleterious to subsequent task performance.
	This verifies that ToI is a more desirable performance metric for task-oriented communication.
	It captures the timing mismatch between obtained status updates in relevance to real-time status at information source.
	The same observation holds for the other two ToI metrics derived in this paper.

	Next,
	we demonstrate that our derived metric captures the end-to-end correlation among three control variables for computation offloading,
	including task generation, transmission and execution.
	Fig.~\ref{AoI_different_utuc} shows the trend of timeliness of obtained status updates under different transmission rates as task generation rate increases, under fixed computation rate.
	As shown in the figure,
	the best achievable ToI performance (minimum points at each curve) gets better as transmission rate increases,
	benefiting from less congestion in the transmission queue.
	However, as the transmission and computation rates have comparable magnitudes,
	the system time of a offloaded task in tandem queue is largely affected by the waiting time in computation queue,
	and thus the ToI performance has marginal improvements as further increase of transmission rate.
	It shows that the best achievable ToI performance is decided by the worse between transmission and computation.
	As for a given combination of available resources,
	the task generation rate dominates the system performance.
	Moreover,
	the optimal task generation rate varies under different combinations.
	It verifies the necessity of factoring in the task generation process during performance metrics design,
	which is usually ignored in conventional delay or throughout metrics.
	In our derived ToI metrics, 
	the dependency of task generation, transmission and computation variables are explicitly characterized.

	Fig.~\ref{AoI_awareness} shows the trend of timeliness of obtained status updates under different task generation policies as transmission rate increases, under fixed computation rate.
	Based on Proposition 1,
	we calculate the optimal task generation rate given values of $\mu_t$ and $\mu_c$,
	and plot the best achievable ToI as stared line.
	As shown in the figure,
	the achievable performance under different task generation rates (2, 3, and 4) converge to the optimal rate policy at different points.
	The zero-wait policy achieves a better performance than optimal rate policy when transmission rate is small,
	benefiting from completely eliminating the waiting time in transmission queue by generating a task only when the channel is idle.
	As the transmission rate increases,
	the performance gain of zero-wait policy decreases, due to less congestion in transmission queue.
	When transmission rate is comparable to the fixed computation rate and further increases,
	the value of ToI grows exponentially under zero-wait policy,
	due to the fact that generated tasks are backlogged in computation queue and thus end-to-end system time is dominated by the waiting time in overloaded computation stage.
	It verifies the necessity of end-to-end system design for computation offloading.

	Fig.~\ref{compare} shows the trend of process-related ToI under different parameters of monitored process.
	As shown in the figure, 
	given a fixed computation rate,
	there exists an optimal combination of task generation and transmission rates.
	Compared with ToI (plotted in dashed line here as a benchmark), 
	the process-related ToI curves are flatter. 
	The reason is that the statistical properties of monitored process is employed in the derivation of process-related ToI expression, 
	which yields a more specific measure of the real-time reconstruction error compared with ToI (simply assumes the error increases linearly with time).
	It shows that by leveraging structure of monitored process, 
	the	real-time reconstruction performance at central controller is more stable as the system load increases. 
	Moreover, 
	the parameters of monitored process has a profound impact on the achievable ToI performance. 
	As the value of $\sigma$ increases,
	the ToI curve becomes more sensitive to the variation of task generation rate.
	It is because that when the monitored process becomes noisy, 
	out-of-date status updates are of less value in predicting the current status at source.

	\section{Timeliness of Information at Fog Tier\\
		M/M/1 -- M/M/1 with Multiple Sources}
	\label{sec:IV:fogTier}
	
	In this section,
	we analyze the timeliness of information for computation offloading at fog tier,
	where the tasks outsourced by edge servers are executed sequentially.
	Task generation at each device is Poisson with rate $\lambda$.
	The service times of transmission and computation are exponential random variables with rate $u_t$ and $u_c$, respectively.
	Here, $u_t$ consists of delay of both wireless transmission and wired transmission.
	Both queues are of infinite size and nonpreemptive.
	Therefore,
	the computation offloading process at fog can be abstracted as M/M/1 queue followed by M/M/1 queue with multiple sources,
	as shown in Fig.~\ref{fog}.
	
	\begin{figure}[h]
		\raggedright{\includegraphics[scale = 0.28]{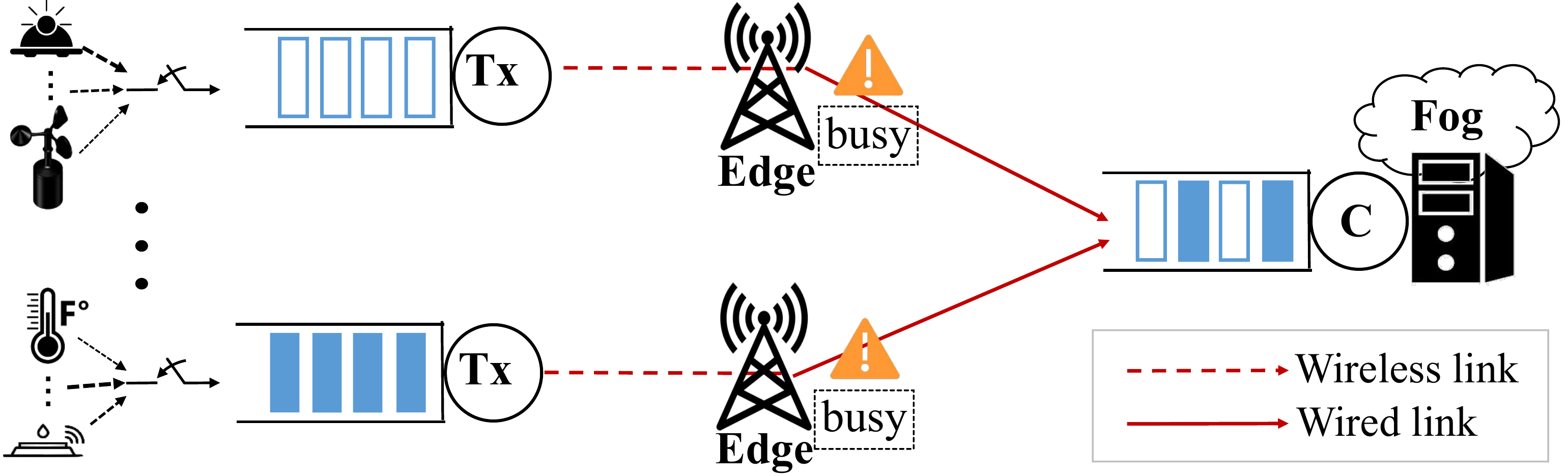}}
		\caption{Queuing model at fog tier.}
		\label{fog}
	\end{figure}
	
	Different from edge servers that are dedicated to a certain group of devices,
	the fog server handles tasks that are randomly outsourced by edge servers at regional level.
	It is hard to obtain prior knowledge about the structure of monitored process.
	In the following, 
	we investigate timeliness of information at fog tier.

	\subsection{Timeliness of Information Analysis}

	Different from the model studied in Sec.~\ref{sec:subsec:edge-agnostic},
	the computation queue at fog tier is abstracted as multi-source M/M/1.
	It captures the key feature that queuing can occur during interarrival times between offloaded tasks from the same device, 
	due to aggregation of tasks from other devices.
	It further complicates the analysis.
	
	Denote $\lambda_i$ as the task generation rate at device $i$,
	and $\lambda_{-i}$ as the aggregated task generation rate from other source devices.
	Since tasks are generated at each device according to a Poisson process,
	the aggregated task generation process is also Poisson.
	In the M/M/1 transmission queue,
	the inter-departure time is statistically identical to the inter-arrival time,
	then the arrival rate of computation queue is $\lambda_i+\lambda_{-i}$.
	Then we derive the timeliness of information for device $i$.
	
	\begin{proposition}
		Assume M/M/1 for transmission queue followed by multi-source M/M/1 for computation queue,
		the timeliness of information is expressed as:
		\begin{equation}\label{}
			\begin{split}
				\bar{\Delta}_{i}
				=&\frac{{\lambda _i^2}}{{\mu _{i,t}^2({\mu _{i,t}} - {\lambda _i})}} + \frac{1}{{{\mu _{i,t}}}} + \frac{1}{{{\mu _c}}}
				+ \frac{{\lambda _i^2}}{{({\mu _c} - \lambda )({\mu _c} - {\lambda _{-i}})}}\\
				&\times \! \!\left[ \frac{1}{{{\mu _c} \!-\! {\lambda _{-i}}}} \!-\! \frac{{{\lambda _i}}}{{{\mu _{i,t}}({\mu _{i,t}} \!-\! {\lambda _i})}}\!+\!\frac{{{\mu _c} \!-\! {\lambda _{-i}}}}{{({\mu _{i,t}} \!-\! {\lambda _i})({\mu _{i,t}} \!+\! {\mu _c} \!-\! \lambda )}}\right]\\
				&{\rm{ + }}\frac{{\lambda _i^2{\lambda _{-i}}}}{{{\mu _c}({\mu _c} - {\lambda _{-i}})}} \! \left[ \frac{2}{{{{({\mu _c} \!-\! {\lambda _{-i}})}^2}}} \!-\! \frac{{{\lambda _i}}}{{{\mu _{i,t}}({\mu _{i,t}} \!-\! {\lambda _i})({\mu _c} \!-\! {\lambda _{-i}})}} \right.\\
				& \left. + \frac{{({\mu _c} - {\lambda _{-i}})}}{{({\mu _{i,t}} - {\lambda _i}){{({\mu _{i,t}} + {\mu _c} - \lambda )}^2}}} \right]
				+ \frac{{{\lambda _i}{\lambda _{-i}}({\mu _c} - \lambda )}}{{{\mu _c}{{({\mu _c} - {\lambda _{-i}})}^2}}} \left[ \frac{1}{{{\lambda _i}}} \right. \\
				&\left. - \frac{{{\lambda _i}}}{{{\mu _{i,t}}({\mu _{i,t}} + {\mu _c} - \lambda )}} + \frac{1}{{{\mu _c} - {\lambda _{-i}}}}\right] + \frac{1}{{{\lambda _i}}}, \nonumber
			\end{split}
		\end{equation}
		where $\lambda = \lambda_i + \lambda_{-i}$.
		
		\begin{proof}
			Following Proposition 1,
			the ToI of a task from device $i$ can be divided as follows:
			\begin{align} \label{eq:multi_average_AoI}
				\bar{\Delta}_i
				=&\lambda_i \bigg( \mathbb{E}[X_{i,n}W_{i,n,t}]+\mathbb{E}[X_{i,n}]\mathbb{E}[S_{i,n,t}]+\mathbb{E}[X_{i,n}W_{i,n,c}] \nonumber \\
				&+\mathbb{E}[X_{i,n}]\mathbb{E}[S_{i,n,c}]
				+\frac{1}{2}\mathbb{E}[X_{i,n}^2] \bigg).
			\end{align}
			
			The terms $\mathbb{E}[X_{i,n}]\mathbb{E}[S_{i,n,t}]$, $\mathbb{E}[X_{i,n}]\mathbb{E}[S_{i,n,c}]$ and $\mathbb{E}[X_{i,n}^2]$ can be obtained based on PDFs of $X_{i,n}$, $S_{i,n,t}$ and $S_{i,n,c}$.
			Since transmission process among multiple devices are independent,
			we evaluate $\mathbb{E}[X_{i,n}W_{i,n,t}]$ following Proposition 1 as:
			\begin{equation}\label{eq:firstterm}
				\mathbb{E}[X_{i,n}W_{i,n,t}]=\frac{\lambda_i}{\mu_{i,t}^2(\mu_{i,t}-\lambda_i)}.
			\end{equation}
			
			As for $\mathbb{E}[X_{i,n}W_{i,n,c}]$,
			the waiting time of the $n$-th task in computation queue depends on the system time of $(n-1)$-th task, and the aggregate other-source tasks.
			Denote $W'_{i,n,c}$ as waiting time caused by $(n-1)$-th task,
			and $W''_{i,n,c}$ as waiting time caused by other-source tasks that arrives within $D_{i,n,t}$:
			\begin{equation}\label{eq:wait}
				W_{i,n,c}=W'_{i,n,c} + W''_{i,n,c}. \nonumber
			\end{equation}
			
			We consider two conditions $I_n$ and $L_n$.
			
			\textbf{(1) \bm{$\mathbb{E}[{X_{i,n}}W_{i,n,c}|I_n]$}:}
			In this case,
			the $(n-1)$-th task is still in the computation queue upon arrival of the $n$-th task.
			Then the sojourn time of $(n-1)$-th task in computation queue is longer than the inter-departure time of the $n$-th task in communication queue ($D_{i,n,t}<T_{i,n-1,c}$).
			Since $T_{i,n-1,c}$ and $D_{i,n,t}$ are independent, following exponential distributions of rate $(\mu_c-\lambda_i-\lambda_{-i})$ and $\lambda_i$, respectively.
			The state probability of $I_n$  is obtained as:
			\begin{equation}
				\begin{split} \label{eq:PIL}
					P(I_n)&=P({D_{i,n,t}} < {T_{i,n-1,c}}) = \frac{{{\lambda _i}}}{{{\mu _c} - {\lambda _{-i}}}}.\\
				\end{split}
			\end{equation}
			
			Since $W'_{i,n,c}>0$, we have:
			\begin{equation} \label{eq:{X_{i,n}}{W_{i,n,c}}|I_n}
				\mathbb{E}[{X_{i,n}}{W_{i,n,c}}|I_n] = \mathbb{E}[{X_{i,n}}W'_{i,n,c}|I_n]+\mathbb{E}[{X_{i,n}}W''_{i,n,c}|I_n]. \nonumber
			\end{equation}
			
			\begin{lemma}
				At state $I_n$, $\mathbb{E}[X_{i,n}W'_{i,n,c}]$ is given as:
				\begin{equation}
					\begin{split} \label{eq:XW1.1}
						\mathbb{E}[X_{i,n}W'_{i,n,c}|I_n]=&\frac{1}{\mu_c-\lambda_i-\lambda_{-i}} \! \left[  \frac{1}{{{\mu _c} - {\lambda _{-i}}}} \!-\! \frac{{{\lambda _i}}}{{{\mu _{i,t}}({\mu _{i,t}} \!-\! {\lambda _i})}} \right.\\
						&\left. +\frac{{{\mu _c} - {\lambda _{-i}}}}{{({\mu _{i,t}} - {\lambda _i})({\mu _{i,t}} + {\mu _c} - \lambda_i-\lambda_{-i} )}}\right]. \nonumber
					\end{split}
				\end{equation}
				
				\begin{proof}
					The proof is given in Appendix D.
				\end{proof}				
			\end{lemma}

			As for $\mathbb{E}[{X_{i,n}}W''_{i,n,c}|I_n]$,
			$W''_{i,n,c}$ depends on the number of other-source tasks arrived during $D_{i,n,t}$,
			as long as the computation time of each individual task.
			Denote $N_{n}$ as the number of other-source tasks,
			$\mathbb{E}[N_n|D_{i,n,t}=y,I_n]=\lambda_{-i}y$.
			We have:
			\begin{equation}\label{}
				\begin{split}
					\mathbb{E}[X_{i,n}W''_{i,n,c}|,D_{i,n,t}=y,I_n]
					=\frac{\lambda_{-i}y}{\mu_c}\mathbb{E}[X_{i,n}|D_{i,n,t}=y]. \nonumber
				\end{split}
			\end{equation}
			
			Then we evaluate $\mathbb{E}[{X_{i,n}}W''_{i,n,c}|I_n]$ as
			\begin{equation}\label{eq:XW2}
				\begin{split}
					&\mathbb{E}[X_{i,n}W''_{i,n,c}|I_n]\\
					=&\int_0^{\infty}\frac{\lambda_{-i}y}{\mu_c}\mathbb{E}[X_{i,n}|D_{i,n,t}=y]f_{D_{i,n,t}|I_n}(y)dy,  \nonumber
				\end{split}
			\end{equation}
			where $\mathbb{E}[X_{i,n}|D_{i,n,t}=y]$ and $f_{D_{i,n,t}|I_n}(y)$ are given in Appendix D.
			
			\textbf{(2) \bm{$\mathbb{E}[{X_{i,n}}W_{i,n,c}|L_n]$}:}
			In this case,
			the $(n-1)$-th task has been executed and left computation queue upon arrival of the $n$-th task.
			Then system of $(n-1)$-th task in computation queue is shorter that the inter-departure time of the $n$-th task in communication queue ($D_{i,n,t}>T_{i,n-1,c}$).
			The state probability of $L_n$  is obtained as:
			\begin{equation}
				\begin{split} \label{eq:PIL2}
					P(L_n)&=P({D_{i,n,t}} \geq {T_{i,n - 1,c}}) = \frac{{{\mu _c} - \lambda_i - \lambda_{-i} }}{{{\mu _c} - {\lambda _{-i}}}},
				\end{split}
			\end{equation}

			Since $W'_{i,n,c}=0$, we have:
			\begin{equation}
				\begin{split} \label{eq:XW|L}
					\mathbb{E}[{X_{i,n}}{W_{i,n,c}}|L_n]
					=\mathbb{E}[{X_{i,n}}|L_n]\mathbb{E}[W''_{i,n,c}|L_n],
				\end{split}
			\end{equation}
			
			We evaluated $\mathbb{E}[X_{i,n}|L_n]$ as:
			\begin{equation}
				\begin{split} \label{eq:X|L}
					\hspace{-1.65mm} \mathbb{E}[X_{i,n}|L_n]=&\int_0^{\infty}\mathbb{E}[X_{i,n}|D_{i,n,t} = y]f_{D_{i,n,t}|L_n}(y)dy\\
					=&\frac{1}{{{\lambda _i}}} \!+\! \frac{1}{{{\mu _c} \!-\! {\lambda _{-i}}}} \!-\! \frac{{{\lambda _i}}}{{{\mu _{i,t}}({\mu _{i,t}} \!+\! {\mu _c} \!-\! \lambda_i-\lambda_{-i} )}}.
				\end{split}
			\end{equation}
			
			As for $W''_{i,n,c}$ ,
			denote $M_{n}$ as the number of backlogged other-source tasks upon the arrival of the $n$-th task.
			Since $D_{i,n,t}>T_{i,n-1,c}$,
			$M_{n}$ cannot be directly derived based on $D_{i,n,t}$.
			Between departure of the $(n-1)$-th task and the arrival of $n$-th task,
			the computation queue can be treated as a M/M/1 queue with single source of $\lambda_{-i}$.
			Referring to
			\cite{xianxin-proof1},
			the stability of M/M/1 queue yields that $M_{n}$ is statistically identical to the number of backlogged tasks when the $(n-1)$-th task left:
			
	\begin{equation}\label{}
				\begin{split}
					\mathbb{E}[W''_{i,n,c}|L_n]=\frac{\mathbb{E}[M_n|L_n]}{\mu_c}
					=\frac{1}{\mu_c}\int_0^{\infty}\lambda_{-i}yf_{T_{i,n-1,c}|L_n}(y)dy. \nonumber
				\end{split}	
			\end{equation}
			
			Since $T_{i,n-1,c}$ and $D_{i,n,t}$ are independent, we have:
			\begin{equation}\label{}
				\begin{split}
					f_{T_{i,n-1,c}|L_n}(y) &= f_{T_{i,n-1,c}|T_{i,n-1,c} \le D_{i,n,t}}(y) \\
					&= ({\mu _c} - {\lambda _{-i}}){e^{ - ({\mu _c} - {\lambda _{-i}})y}}. \nonumber
				\end{split}
			\end{equation}
			
			Then we have:
			\begin{equation}
				\begin{split} \label{eq:W_L}
					\mathbb{E}[W''_{i,n,c}|L_n]
					=\frac{\lambda_{-i}}{\mu_c(\mu_c-\lambda_{-i})}.
				\end{split}	
			\end{equation}
			
			By putting \eqref{eq:X|L} and \eqref{eq:W_L} in \eqref{eq:XW|L}, the expression for $\mathbb{E}[{X_{i,n}}W_{i,n,c}|L_n]$ is obtained.
			We evaluate $ \mathbb{E}[{X_{i,n}}{W_{i,n,c}}]$ as
			\begin{equation}
				\begin{split} \label{eq:X1W_fenjie}
					\mathbb{E}[{X_{i,n}}{W_{i,n,c}}] =& P(I_n)\mathbb{E}[{X_{i,n}}{W_{i,n,c}}|I_n]\\
					& + P(L_n)\mathbb{E}[{X_{i,n}}{W_{i,n,c}}|L_n].
				\end{split}
			\end{equation}
			
			By putting \eqref{eq:X1W_fenjie} and \eqref{eq:firstterm} in \eqref{eq:multi_average_AoI},
			we can get the final expression for timeliness of information.			
		\end{proof}
	\end{proposition}

	\noindent\textbf{Zero-wait Task Generation:}
	We also evaluate process-related timeliness of information under zero-wait task generate policy.
	
	\begin{corollary}
		Under zero-wait policy,
		the timeliness of information in M/M/1--multi-source M/M/1 is:
		\begin{equation}\label{}
			\begin{split}
				\bar{\Delta}_i^* = \frac{2}{\mu_{i,t}}+\frac{1}{\mu_c}+
				\frac{1}{\mu_c-\lambda_{-i}}\left(\frac{\mu_{i,t}}{\mu_c-\mu_{i,t}-\lambda_{-i}}+\frac{\lambda_{-i}}{\mu_c}\right). \nonumber
			\end{split}
		\end{equation}
		
		\begin{proof}
			In this case, $W_{i,n,t}=0$, $X_{i,n}=S_{i,n-1,t}$, and $D_{i,n-1,t}=S_{i,n-1,t}$. Based on \eqref{eq:multi_average_AoI}, we have:
			\begin{equation}
				\begin{split} \label{eq:age_peaK2}
					\bar{\Delta}_i^* = &\mathbb{E}[S_{i,n,t}]+\mathbb{E}[S_{i,n,c}]+\mathbb{E}[W_{i,n,c}]
					+\frac{1}{2}\frac{\mathbb{E}[S_{i,n,t}^2]}{\mathbb{E}[S_{i,n,t}]},
				\end{split}
			\end{equation}
			where $\mathbb{E}[S_{i,n,t}]$, $\mathbb{E}[S_{i,n,c}]$ and $\mathbb{E}[S_{i,n,t}^2]$ can be derived based on PDFs of $S_{i,n,t}$ and $S_{i,n,c}$.
			
			As for $\mathbb{E}[W_{i,n,c}]$,
			we also consider two conditions:
			\begin{equation}
				\begin{split} \label{}
					\mathbb{E}[W_{i,n,c}] = P(I_n)\mathbb{E}[{W_{i,n,c}}|I_n] + P(L_n)\mathbb{E}[{W_{i,n,c}}|L_n], \nonumber
				\end{split}
			\end{equation}
			where $P(I_n)$, $P(L_n)$ and $\mathbb{E}[{W_{i,n,c}}|L_n]$ can be obtained from \eqref{eq:PIL},  \eqref{eq:PIL2} and \eqref{eq:W_L} by substituting $\lambda_i$ with $\mu_{i,t}$, respectively.
			
			We evaluate $\mathbb{E}[{W_{i,n,c}}|I_n]$ as:
			\begin{equation}
				\begin{split} \label{}
					\mathbb{E}[{W_{i,n,c}}|I_n] = \mathbb{E}[{W'_{i,n,c}}|I_n] + \mathbb{E}[{W''_{i,n,c}}|I_n], \nonumber
				\end{split}
			\end{equation}
			where $\mathbb{E}[{W'_{i,n,c}}|I_n] $ can be derived as \eqref{eq:XW1.2}.
			As before, we evaluate $\mathbb{E}[{W''_{i,n,c}}|I_n]$ as:
			\begin{equation}\label{eq:last}
				\begin{split}
					\mathbb{E}[{W''_{i,n,c}}|I_n] &=\int_0^\infty \mathbb{E}[{W''_{i,n,c}}|I_n,D_{i,n,t}=y]f_{D_{i,n,t}|I_n}(y)dy \\
					&=\frac{\lambda_{-i}}{\mu_{c}(\mu_{c}-\lambda_{-i})}.  \nonumber
				\end{split}
			\end{equation}
		
			Therefore, the expression for \eqref{eq:age_peaK2} can be obtained.			
		\end{proof}
	\end{corollary}
		
	\subsection{Numerical Results and Analysis}
	In this section,
	we present numerical results to demonstrate the impact of system parameters on timeliness of information at fog tier,
	including task load from other devices, task generation rate, transmission rate, and computation rate.
	We also present the achievable performance under zero-wait task generation policy.
	
	\begin{figure*}
		\begin{minipage}{0.33\textwidth}
			\raggedright
			\includegraphics[width=\textwidth]{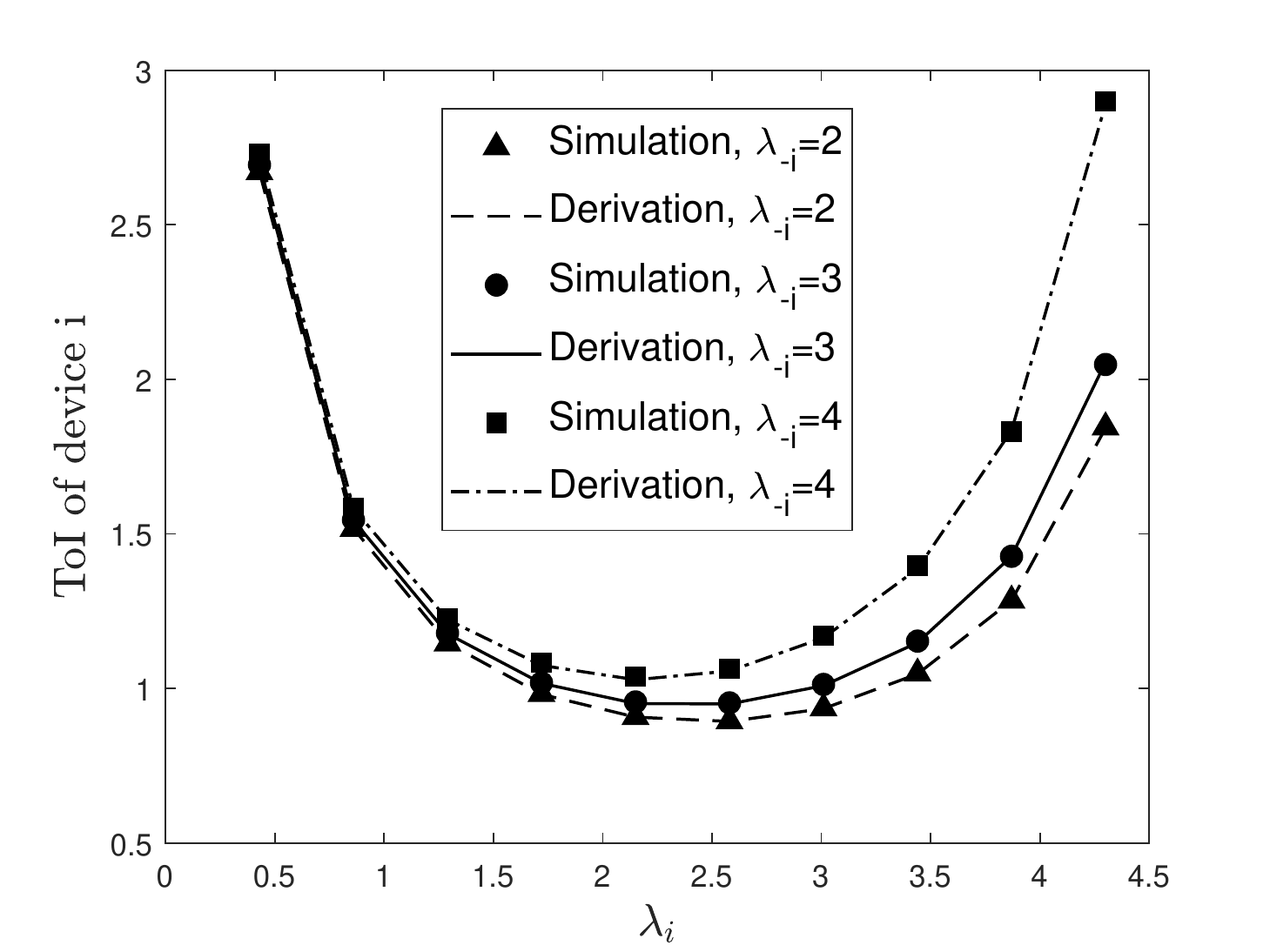}
			\caption{ToI of device i as task generation rate increases under different system load. $\mu_{i,t}=5,\mu_c=9$.}
			\label{multi_aoi_monte}
		\end{minipage}
		\hfill
		\begin{minipage}{0.33\textwidth}
			\centering
			\includegraphics[width=\textwidth]{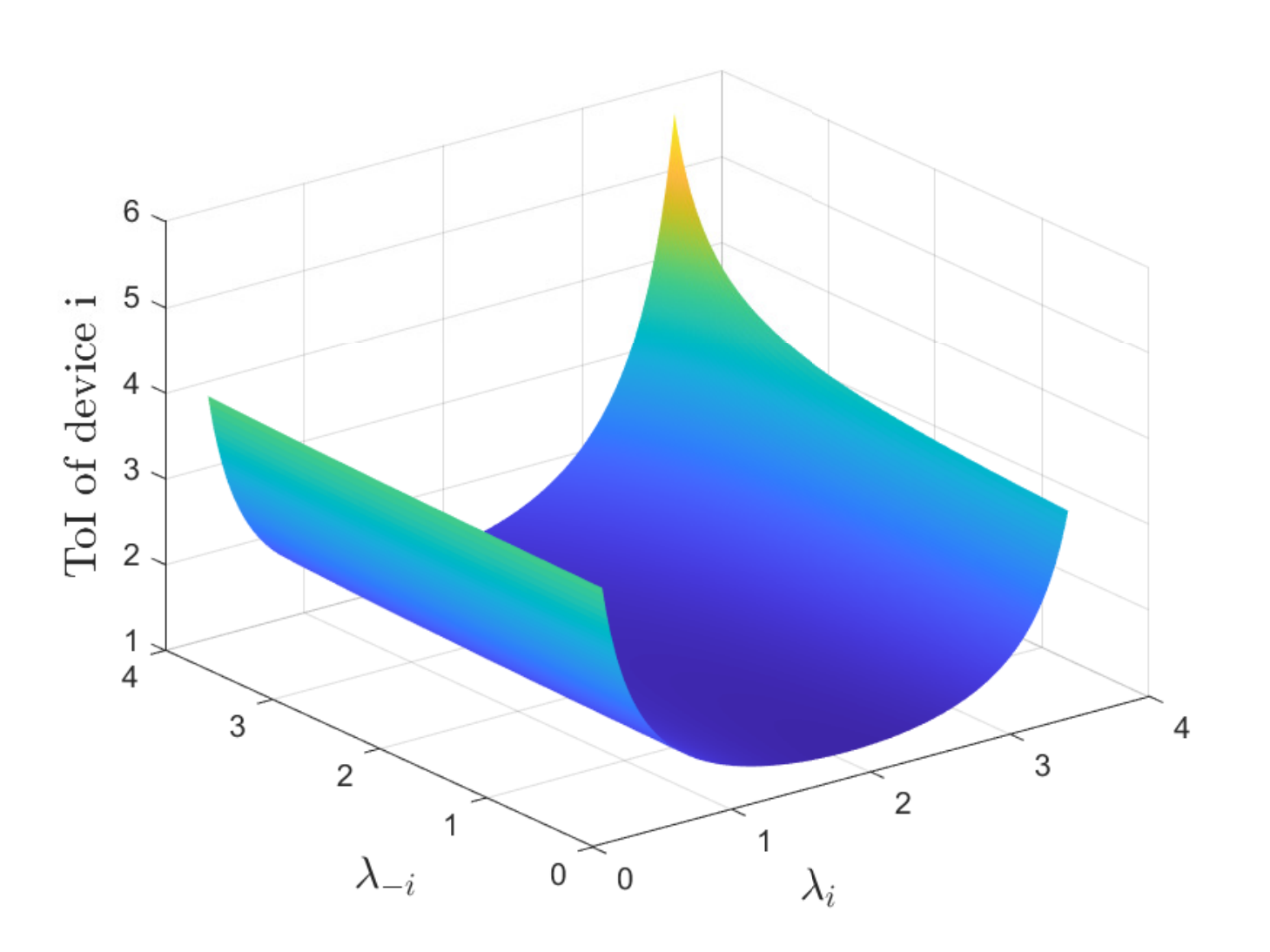}
			\caption{ToI of device $i$ as task generation rate at device $i$ and other devices increases. $\mu_t=4,\mu_c=8$.}
			\label{AoI_multi}
		\end{minipage}
		\hfill
		\begin{minipage}{0.32\textwidth}
			\raggedleft
			\includegraphics[width=\textwidth]{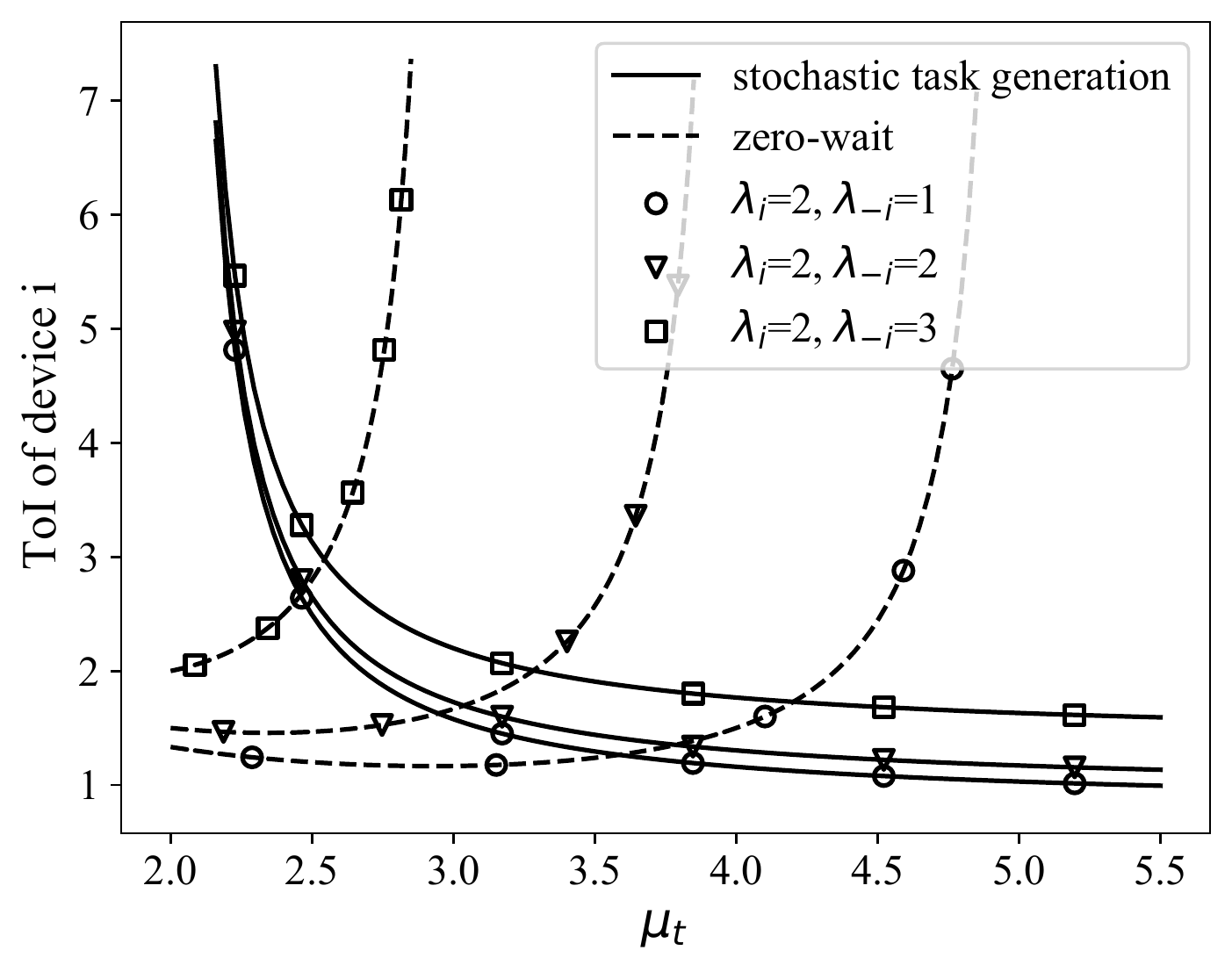}
			\caption{ToI of device $i$ under different task generation policies as transmission rate increases. $\mu_c=6$.}
			\label{multi_AoI_awareness}
		\end{minipage}
	\end{figure*}
	
	First, we compare the results of our analysis with Monte Carlo simulations, generating $N=10^6$ tasks and computing ToI and for each. 
	As shown in Fig.~\ref{multi_aoi_monte}, 
	the simulation results perfectly match the theoretically derived curves,
	verifying the soundness of our derivation.
	
	Fig.~\ref{AoI_multi} shows the trend of ToI of device $i$ under different task generation rates, as the aggregate task load from other devices increases.
	The transmission rate is set as 4 and the computation rate is set as 8.
	As shown in the figure,
	when task generation rate at device $i$ is small (along the axis of $\lambda_i <1$),
	the increasing task load from other devices has marginal impact on the ToI performance of device $i$.
	This is because that the obsoleteness of device $i$'s status updates is mainly caused by ``lazy'' task generation,
	thus the interarrival time dominates ToI for device $i$.
	As $\lambda_i$ increases,
	the impact of task load from other devices becomes more profound,
	which leads to obvious increment along the axis of $\lambda_i >3$,
	where tasks generated by other devices congest the shared computation queue.

	Fig.~\ref{multi_AoI_awareness} shows the trend of ToI of device $i$ under different task generation policies as transmission rate increases.
	The task generation rate of device $i$ is set as 2,
	while the aggregate task generation rate at other devices increases from 1 to 3.
	As for the same combination of task generation rates (e.g., $\lambda_i=2, \lambda_{-i}=1$),
	zero-wait policy achieves a better performance compared with stochastic task generation under low transmission rate regime ($u_t<3.6$),
	benefiting from eliminating the waiting time in transmission queue.
	The performance gain becomes smaller as the increasing transmission rate.
	When $u_t>3.6$,
	the value of ToI grows exponentially under zero-wait policy,
	due to the fact that the system time is dominated by waiting time in computation queue caused by increasing task generation rates at both device $i$ and other devices.
	Note that as the aggregate task generation rate at other devices increases,
	the benefits of adopting zero-wait policy at device $i$ is weakened.

	\section{An Illustrative Case Study}
	\label{sec:V:caseStudy}
	The closed-form expressions derived in Sec.~\ref{sec:III:edgeTier} and Sec.~\ref{sec:IV:fogTier} provide explicit dependencies among task generate rate at device, transmission rate and computation rate.
	The derived expressions can be employed as performance metrics for computation offloading strategy design.
	In this section,
	we employ ToI derived in Sec.~\ref{sec:subsec:edge-agnostic} as an illustrative example,
	and investigate the computation offloading scheme at edge tier.
	For a set of IoT devices, our objective is to minimize the timeliness of status updates extracted from computation task flows by optimizing variables of task generation, bandwidth allocation, and computation resource allocation.
	
	\subsection{Problem Formulation}
	Consider a MEC-assisted networked control system, which consists of an edge server and a set of $\mathcal{M}$ IoT devices,
	where $M = |\mathcal M|$ is the number of devices.
	The devices are deployed at various monitoring spots to obtain timely situational awareness by continuously generating and offloading perception tasks to edge server to extract status updates.
	Each device generates tasks as a Poisson process with rate $\lambda_m$.
	As for task offloading,
	we assume an OFDMA-based system where each device has a pre-assigned sub-channel for task offloading \cite{yang-iot}.
	The offloaded tasks from multiple users are executed at edge server in parallel mode with virtualization,
	where computation resource at edge server is reserved among devices.
	
	As for task offloading,
	assume that each device can perfectly estimate its own local channel state information by using downlink pilot signals. 	
	For simplicity, 
	we consider a quasi-static scenario where uplink channels remain unchanged over the time period of observation\cite{channelmodel}.
	Each device offloads task with constant transmission power $p_m$.
	Denote $\beta_m$ as the proportion of bandwidth allocated to device $m$,
	then the transmission rate at device $m$ can be obtained as:
	\begin{equation}\label{Rm}
		R_m = B\beta_mlog_2(1+\frac{p_mh_m^2}{N_0})\; ,
		(m\in \mathcal M) \; .
	\end{equation}
	where $B$ denotes the total bandwidth,
	$N_0$ denotes Gaussian noise,
	$h_m$ denotes the channel propagation coefficient.
	
	Assume that task size $D_m$ varies over time following exponential distribution,
	denote $\mu^t_m$ as task transmission rate at device $m$,
	then we have:
	\begin{equation}\label{umt}
		\mu^t_m = \mathbb{E}[\frac{R_m}{D_m}]\; ,
		(m\in \mathcal M) \; .
	\end{equation}
	
	The total bandwidth allocated to all devices cannot exceed the total available bandwidth, then we have:
	\begin{equation}\label{constraints:transShare}
		\sum\limits_{m=1}^M \beta_m \leq 1\; ,
		(m\in \mathcal M) \; .
	\end{equation}
	
	As for task execution,
	denote  $f_m$ as the computation resources allocated to device $m$,
	$c_m$ as the number of CPU cycles required to process one bit.
	Then the task execution rate for device $m$ can be obtained as:
	\begin{equation}\label{umc}
		\mu^c_m = \mathbb{E}[\frac{f_m}{c_mD_m}]\; ,
		(m\in \mathcal M) \; .
	\end{equation}
	
	The total computation resources allocated to all devices cannot exceed the computation capability at edge server:
	\begin{equation}\label{constraints:comShare}
		\sum\limits_{m=1}^M f_m \leq f_{max}\; ,
		(m\in \mathcal M) \; .
	\end{equation}
	
	The timeliness of status updates obtained at device $m$ can be characterized as derived in expression \eqref{eq:single average age}:
	\begin{equation}\label{Am}
		\begin{split}
			\Delta_m=&\frac{{{(\lambda_m)^2}}}{{(\mu^t_m)^2({\mu^t_m} - \lambda_m )}} + \frac{1}{{{\mu^t_m}}} + \frac{{{(\lambda_m)^2}}}{{(\mu^c_m)^2({\mu_m^c} - \lambda_m )}} \\
			& +\frac{1}{{{\mu^c_m}}} + \frac{{{(\lambda_m)^2}}}{{{\mu_m^t}{\mu^c_m}({\mu^c_m} + {\mu^c_m} - \lambda_m )}} + \frac{1}{\lambda_m}. 
		\end{split}
	\end{equation}
	
	We are interested in minimizing the steady-state timeliness of information among a set of IoT devices under transmission and computation resource constraints.
	To ensure user fairness, 
	we aim to minimize the maximum ToI among the devices. 
	We introduce an auxiliary variable $\tau$ with $\Delta_m \leq \tau$. 
	 Then the Min-Max programming problem can be formulated as a minimization problem as follows:
	\begin{align} \label{}
		\textbf{(P1)} \min\limits_{\{\lambda_m,\beta_m,f_m\}_{m=1}^M} \ \ &\tau\nonumber \\
		s.t. \quad\quad\  &\eqref{Rm}-\eqref{Am}    \nonumber      \\
		&\Delta_m \leq \tau,  \quad \forall m \in \mathcal{M} \label{tau}  \\
		&\lambda_m\leq\mu^t_m,  \quad \forall m \in \mathcal{M}  \label{lt} \\
		&\lambda_m \leq\mu^c_m,  \quad \forall m \in \mathcal{M}  \label{lc} \\
		&\lambda_m \geq 0, \beta_m \geq 0, f_m \geq 0  \quad \forall m \in \mathcal{M}  \nonumber
	\end{align}
	where constraints \eqref{lt} and \eqref{lc} guarantee the stability of transmission and computation queues.
	In this formulation,
	$\lambda_m$, $\beta_m$, and $f_m$ are continuous variables.
	While the formulated problem incorporates mostly linear constraints,
	solving it is nontrivial due to the coupling among variables in constraints \eqref{Am}.

	\subsection{Algorithm Design}
	We observe that the formulated problem is a multi-convex problem (see Appendix E), 
		thus we propose a solution procedure based on Proximal Block Coordinate Descent\cite{xianxin-proof3}.
	
	At each iteration $i$ of the proposed solution,
	the original problem is decomposed into two convex sub-problems and solved iteratively.
	Given the pre-fixed values of $\{\beta_m^{(i)}, f_m^{(i)}\}_{m=1}^M$,
	the task generation rate at each device $\{\lambda_m^{(i)}\}_{m=1}^M$ can be obtained by solving the following subproblem using tools for disciplined convex programming.
	\begin{align} \label{}
		\textbf{(P2)} \min\limits_{\{\lambda_m\}_{m=1}^M} \ \ &\tau \nonumber \\
		s.t. \quad\quad\
		& \eqref{Rm},\eqref{umt},\eqref{umc}, \eqref{Am}-\eqref{lc}   \nonumber  \\
		&\lambda_m \geq 0, \quad \forall m \in \mathcal{M}  \nonumber
	\end{align}
	
	Then the values of $\{\beta_m^{(i+1)}, f_m^{(i+1)}\}_{m=1}^N$ can be updated by solving the following subproblem using tools for disciplined convex programming.
	\begin{align} \label{}
		\textbf{(P3)} \min\limits_{\{\beta_m,f_m\}_{m=1}^M} \ \ &\tau \nonumber \\
		s.t. \quad\quad\
		&\eqref{Rm}-\eqref{lc}  \nonumber    \\
		&\beta_m \geq 0,f_m \geq 0, \quad \forall m \in \mathcal{M}  \nonumber
	\end{align}
	The procedure iterates until values of $\lambda_m, \beta_m, f_m$ converge.

	\subsection{Performance Evaluation}
	In this section,
	we present simulation results to demonstrate the performance of our proposed strategy.
	The simulation parameters are set as follows.
	IoT devices are randomly distributed in the coverage area of a base station with a radius of $200m$.
	Each device continuously generates computation tasks following Poisson process.
	Assume Rayleigh fading channel models as specified in \cite{channelmodel}.
	The total bandwidth is set as $10\rm KHz$, the transmission power at device is set as $250\rm mW$,
	the computation capability $f_{max}=5\rm GHz$ while $c_m=30\rm cycle/bit$.
	The task size follows exponential distribution with $\bar{D}_m\in [50,300]$ bits.
	
	To demonstrate the benefit of jointly optimizing task generation, bandwidth allocation and computation resource allocation using our derived ToI metric,
	we compare the performance of our proposed solution with six other design principles:
	\begin{itemize}
		\item \emph{uniform computation allocation}: the task generation rate and bandwidth allocation is jointly optimized as proposed, while computation resource is evenly allocated among devices.  
		\item \emph{proportional communication allocation}: the task generation rate and computation resource allocation is jointly optimized as proposed, while bandwidth is allocated among devices reversely proportional to channel condition\cite{HFEL}.
		\item \emph{proportional-uniform resource allocation}: the task generation rate is optimized as proposed, given proportionally allocated bandwidth and evenly allocated computation resource among devices.   	
		\item \emph{task-aware computation allocation} : the task generation rate and bandwidth allocation is jointly optimized as proposed, while computation resource is allocated as in \cite{policy1}, where $f_m = \frac{\sqrt{\alpha_m\bar{D}_m}}{\sum_{m=1}^{M}\sqrt{\alpha_m\bar{D}_m}}f_{max}$.
		\item \emph{task and channel-aware resource allocation}: the task generation rate is optimized as proposed, while computation resource is allocated as in  \cite{policy1}, and the bandwidth is allocated as in \cite{policy2}, where $\beta_m = \frac{\sqrt{\bar{D}_m/R_m}}{\sum_{m=1}^{M}\sqrt{\bar{D}_m/R_m}}$.
		\item \emph{fixed task generation}: the bandwidth and computation resource allocation is jointly optimized as proposed, fixing the task generation rates at all devices as 0.1. 
	\end{itemize} 
	
		\begin{figure}[h]
		\centerline{\includegraphics[scale = 0.6]{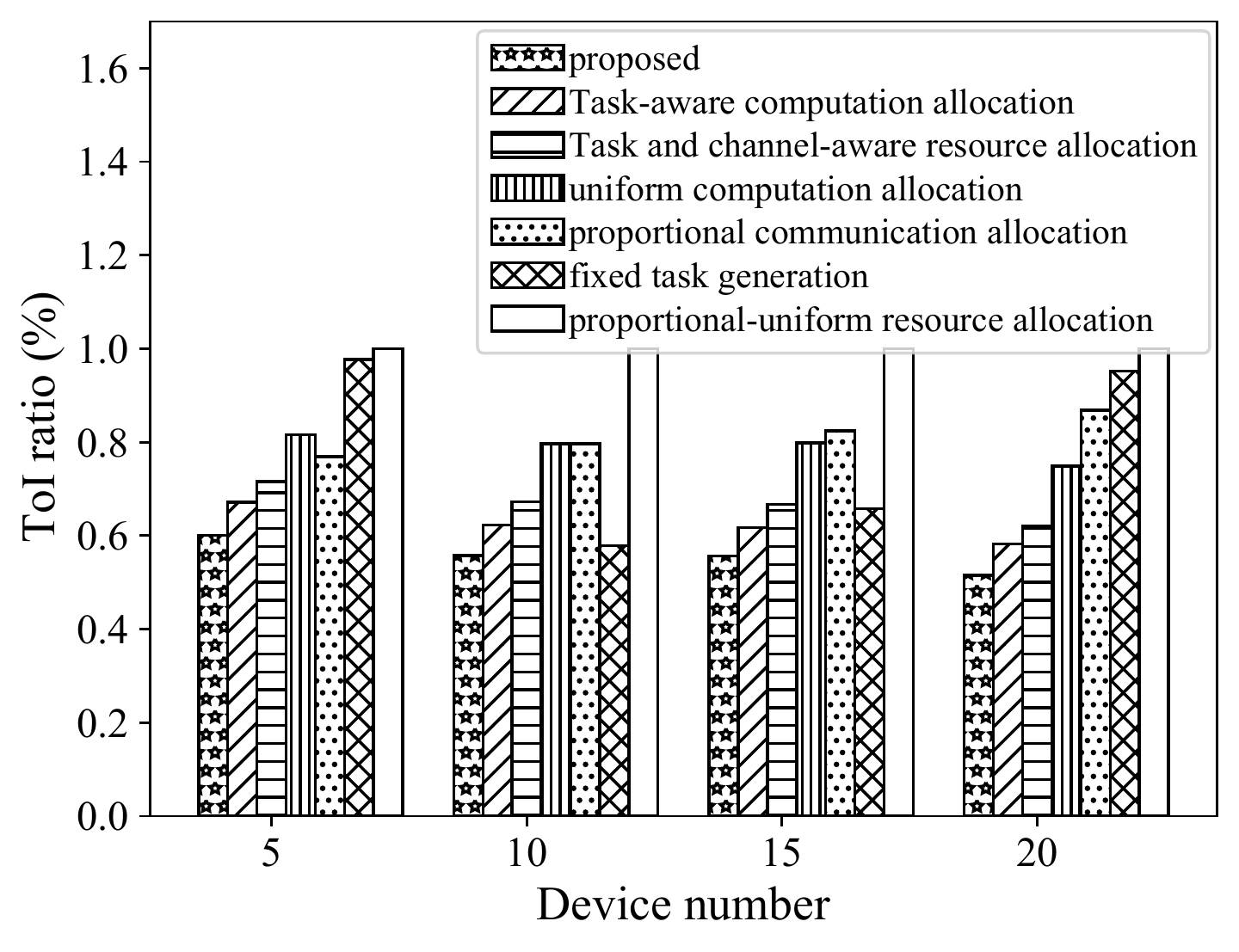}}
		\caption{ToI performance as number of devices increases.}
		\label{policy_compare}
	\end{figure}

	\begin{figure*}[t!]
		\begin{subfigure}[t]{0.33\textwidth}
			\raggedright
			\includegraphics[width=\textwidth]{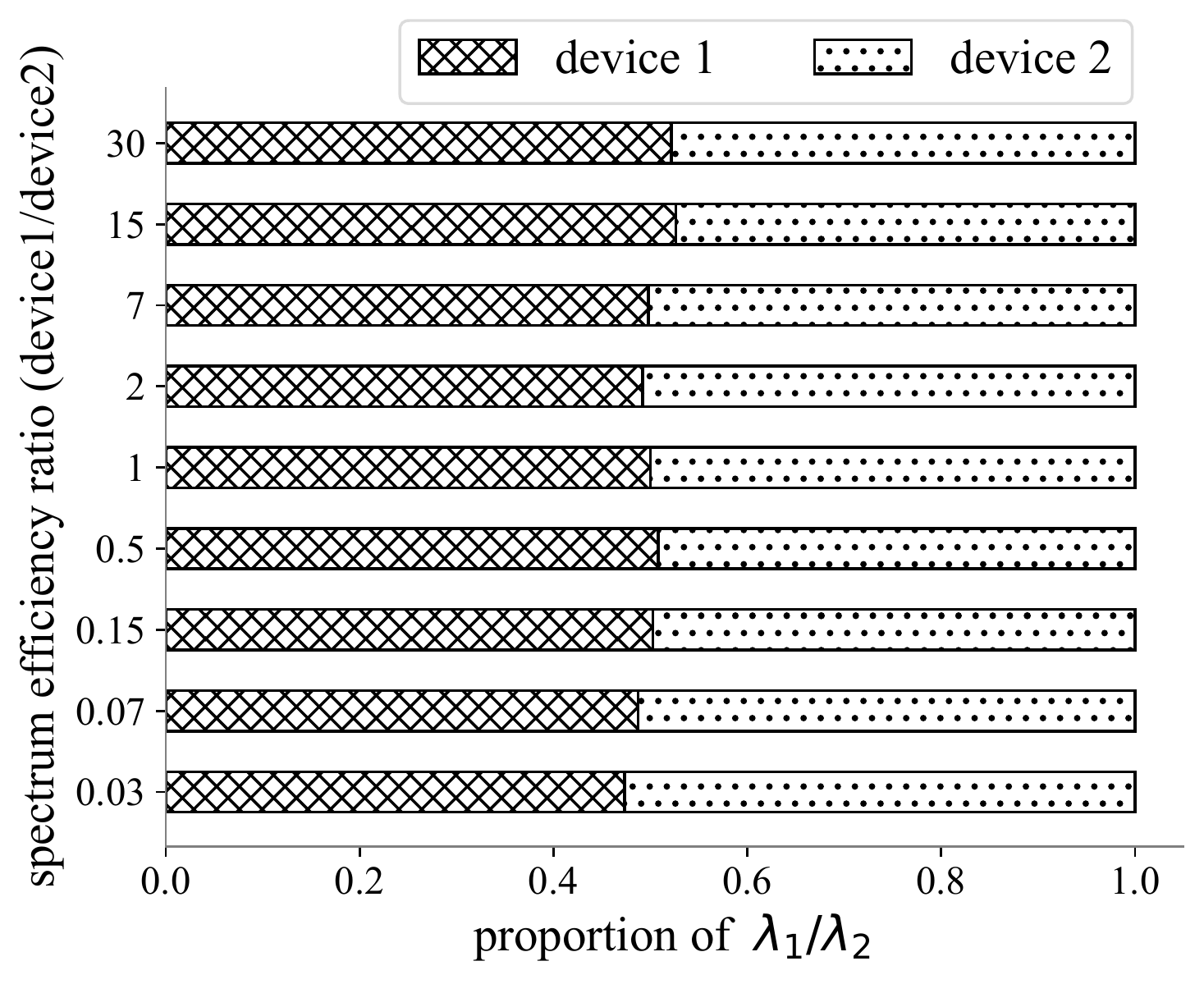}
			\caption{Task generate rate \centering}
			\label{proportion_lambda}
		\end{subfigure}
		\begin{subfigure}[t]{0.33\textwidth}
			\centering
			\includegraphics[width=\textwidth]{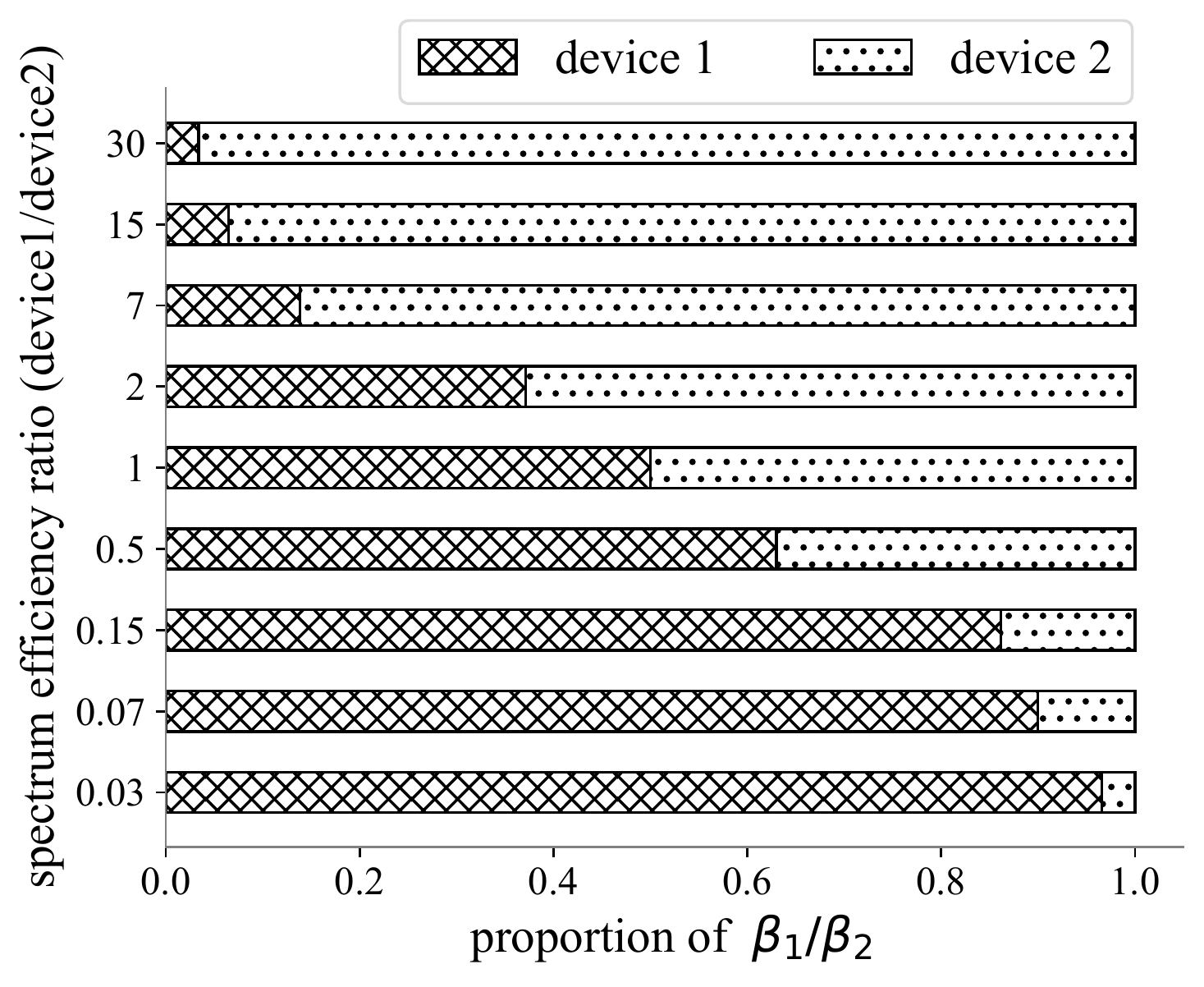}
			\caption{Bandwidth allocation \centering}
			\label{proportion_beta}
		\end{subfigure}
		\begin{subfigure}[t]{0.33\textwidth}
			\centering
			\includegraphics[width=\textwidth]{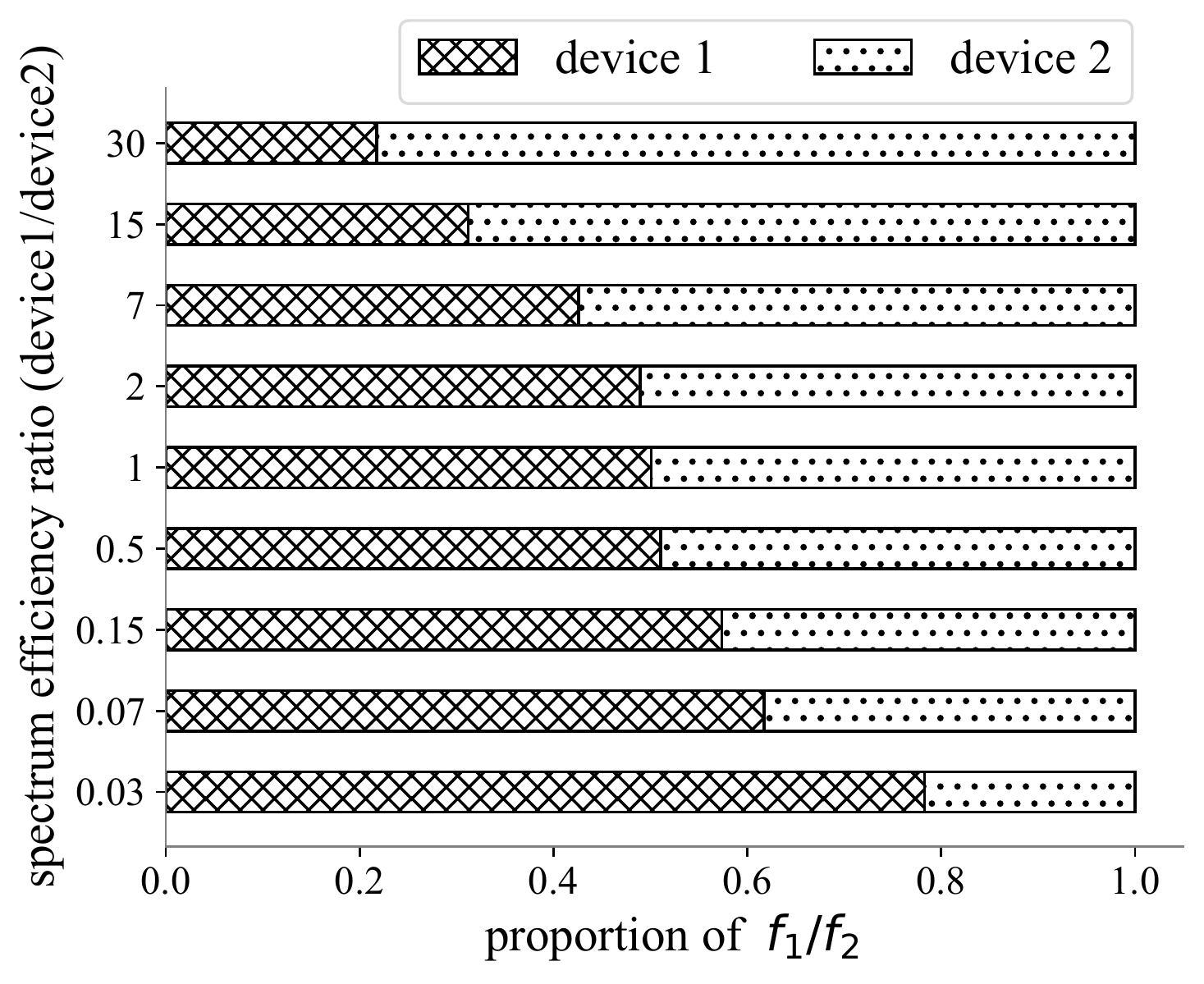}
			\caption{Computation resource allocation \centering}
			\label{proportion_f}
		\end{subfigure}
		\caption{An illustrative example of resource allocation between two users under our proposed strategy.}
		\label{proportion}
	\end{figure*}

	Fig.~\ref{policy_compare} shows the performance of ToI for status updates obtained under different strategies, as the number of devices increases.
	For better illustration, 
	we adopt the ratio of ToI performance regarding ``proportional-uniform resource allocation'' strategy as Y-axis.
	As shown in the figure,
	our proposed strategy always yields the best performance in terms of obtaining status updates in a timely manner.
	Among the other six strategies,
	the performance of ``proportional-uniform resource allocation'' strategy is always the worst. 
	It shows that a naive resource allocation strategy cannot fully exploit the system resources, 
	and thus verifies the necessity of a problem tailored computation offloading strategy.

	Moreover, 
	the  performance of``fixed task generation'' is significantly worse when the number of devices is set as 5 and 20.
	It shows that the interarrival time at transmission queue has a profound impact on system time in either case when the system is light-loaded or crowded.
	In such cases, only optimizing resource allocation among devices is not enough, 
	the information generation policy should also be factored in system design.
	This verify the necessity of a rigorous derivation of the dependency among task generation, transmission and computation variables.
	Note that since ``task-aware computation allocation''  and ``task and channel-aware resource allocation'' strategies takes the size of offloaded task into consideration while allocating system resources, 
	they always yield better performance compared with ``uniform computation allocation'' and ``proportional communication allocation'' .
	However, 
    such  \emph{one-shot} optimization manner cannot capture the temporal correlation among consecutively generated tasks, 
    and thus our proposed strategy outperforms these strategies.

	Fig.~\ref{proportion} shows an illustrative example of our proposed strategy.
	We randomly choose two users, and adjust their spectrum efficiency ratio based on constraints \eqref{Rm}.
	As shown in Fig.~\ref{proportion}(b),
	more bandwidth is always allocated to device with worse channel to compensate the difference in transmission rates.
	Moreover,
	as shown in Fig.~\ref{proportion}(c),
	in case when the difference of channel condition is relatively large,
	more computation resources are allocated to the device with worse channel to even out the difference of system time in transmission queue.
	Note that as shown in Fig.~\ref{proportion}(a), 
	the task generation rate remains relatively the same for two users under all circumstances.
	It shows that our proposed solution achieves timely computation offloading while ensuring user fairness.

	\section{Conclusion}
	\label{sec:VI:conclusion}
To support task-oriented computation offloading in multi-tier computing network,  
it is essential to factor in the context of information into system design philosophy. 
In this sense, 
we abstracted the computation offloading at edge tier and fog tier as two-stage tandem queues, 
and derived the closed-form expressions for timeliness of information. 	
Moreover, 
we exploited the structure of Gauss-Markov process to model process-related timeliness of information. 
The obtained expressions capture the dependency among task generation, task transmission and task execution, along with their impact on the timeliness of obtained status updates. 
Based on the theoretical results, 
we developed a computation offloading optimization problem at edge tier, 
where the timeliness of status updates is minimized among multiple devices by joint optimization of task generation, bandwidth allocation, and computation resource allocation. 
Simulation results demonstrated the dependency among system parameters and the effectiveness of our proposed strategy.
The combination of metrics developed in this paper could be further exploited to design and analyze computation offloading strategies in various scenarios, 
including coexistence of devices monitoring different physical processes, resource provisioning in multi-tier computing networks, etc.

	\appendices
	\section{Proof of Lemma 2}
	Denote $B_{n,t}$ as the inter-departure time of transmission queue, $D_{n,t}=S_{n,t}+B_{n,t}$.
	since service time $S_{n,t}$ are independent with $B_{n,t}$, $W_{n,t}$ and $X_n$,
	we evaluate conditioned PDF of $D_{n,t}$ as 
	\begin{equation}
		\begin{split}\label{eq:S juanji B}
			&{f_{{D_{n,t}}|{W_{n,t}}= 0,{X_n} = x}}(t)\\
			= &{f_{{S_{n,t}}}}(t)\otimes {f_{{B_{n,t}}|{W_{n,t}}= 0,{X_n} = x}}(t),
		\end{split}
	\end{equation}
	where $f_{S_{n,t}}(t)=\mu_t e^{-\mu_t t}$.
	
	Next, we evaluate the conditioned PDF of $B_{n,t}$.
	Since $B_{n,t}=x-T_{n-1,t}$, conditioned PDF of $B_{n,t}$ is
	\begin{equation}\label{eq:fB|0x}
		\begin{split}
			{f_{{B_{n,t}}|{W_{n,t}} = 0,{X_n} = x}}(t) 
			=& f_{T_{n-1,t}|W_{n,t}=0,X_{n}=x}(x-t) \\
			=& \frac{1}{{{P_{idle,t}}}}{f_{{T_{n-1,t}}}}(x - t).
		\end{split}	
	\end{equation}
	
	Combined with \eqref{eq:S juanji B} and \eqref{eq:fB|0x}, the conditioned PDF of $D_{n,t}$ is obtained.

	\section{Proof of Lemma 3}
	Introduce inter-departure time $D_{n,t}$ as an intermediate variable, we have:
	\begin{align}\label{wt+wc}
		\hspace{-8mm} &{f_{{W_{n,t}} + {W_{n,c}}|W_{n,c} > 0,{W_{n,t}} = 0,{X_n} = x,{S_{n,t}} = s}}(t)  \nonumber \\
		=& f_{{W_{n,c}}|{W_{n,t}} = 0,{X_n} = x,{S_{n,t}} = s}(t) ~~(t>0)  \\
		=& \int_s^{s + x} \!\!\! {{f_{{D_{n,t}}|{W_{n,t}} = 0,{X_n} = x,{S_{n,t}} = s}}} (y){f_{{W_{n,c}}|{D_{n,t}} = y}}(t)dy ~(t\!>\!0), \nonumber
	\end{align}
	where the latter is known as \eqref{eq:W-S}.
	
	As for the former, since $B_{n,t}$ and $S_{n,t}$ are independent, based on \eqref{eq:fB|0x}, we have:
	\begin{equation}
		\begin{split}\label{eq:fB|0xs}
			{f_{{B_{n,t}}|{W_{n,t}} = 0,{X_n} = x, S_{n,t}=s}}(t)
			&=\frac{1}{{{P_{idle,t}}}}{f_{{T_{n-1,t}}}}(x - t).
		\end{split}
	\end{equation}
	
	Given $W_{n,t}=0$ and $S_{n,t}=s$, $D_{n,t}=B_{n,t}+s$, we have:
	\begin{equation}\label{}
		\begin{split}
			&{f_{{D_{n,t}}|{W_{n,t}} = 0,{X_n} = x,{S_{n,t}} = s}}(t)\\
			=&{f_{{B_{n,t}}|{W_{n,t}} = 0,{X_n} = x,{S_{n,t}} = s}}(t-s)\\
			=& \frac{1}{{{P_{idle,t}}}}{f_{{T_{n-1,t}}}}(x + s - t),~~~~~~s < t \le s + x.
		\end{split}
	\end{equation}
	
	Then the conditional distribution of waiting time in \eqref{wt+wc} is obtained.

	\section{Proof of Lemma  4}
	We consider two conditions:transmission queue is busy or idle, then we have:
	\begin{equation}
		\begin{split}\label{eq:fW_fenjie2}
			&{f_{{W_{n,t}} + {W_{n,c}}|{X_n} = x,{S_{n,t}} = s,W_{n,t}+W_{n,c}>0}}(t) \\
			=&{P_{busy,t}}{f_{{W_{n,t}} + {W_{n,c}}|{W_{n,t}}>0,{X_n} = x,{S_{n,t}} = s}}(t)\\
			&+{P_{idle,t}}{f_{{W_{n,t}} + {W_{n,c}}|{W_{n,c}}>0,{W_{n,t}}=0,{X_n} = x,{S_{n,t}} = s}}(t).
		\end{split}
	\end{equation}

	Since $W_{n,t} + W_{n,c}>0$, if $W_{n,t}>0$, then it can be further divided into two conditions based on $W_{n,c}$:
	\begin{equation}\label{eq:fWt_fenjie}
		\begin{split}
			&{f_{{W_{n,t}} + {W_{n,c}}|{W_{n,t}>0},{X_n} = x,{S_{n,t}} = s}}(t)\\
			=&{P(W_{n,c}=0|W_{n,t}>0,X_n=x,S_{n,t}=s)}\\
			&\times {f_{{W_{n,t}} + {W_{n,c}}|W_{n,c}=0,{W_{n,t}}>0,{X_n} = x,{S_{n,t}} = s}}(t)\\
			&+{P(W_{n,c}>0|W_{n,t}>0,X_n=x,S_{n,t}=s)}\\
			&\times {f_{{W_{n,t}} + {W_{n,c}}|W_{n,c}>0,{W_{n,t}}>0,{X_n} = x,{S_{n,t}} = s}}(t).
		\end{split}
	\end{equation}
	
	For the first term, it refers to the case when sojourn time of \textcolor{blue}{$(n-1)$}-th task in computation queue is less than the service time of $n$-th task in communication queue, that is
	\begin{equation}
		\begin{split} \label{eq:PWc0}
			\hspace{-4mm} P({W_{n,c}} \!\!=\! 0|{W_{n,t}}\! >\! 0,{X_n} \!=\! x,{S_{n,t}} \!=\! s)
			=& \int_0^s \!\!{{f_{{T_{n-1,c}}}}\!(t)dt}  \\
			=& 1 - {e^{ - (\mu_c-\lambda)s}}.
		\end{split}
	\end{equation}
	
	Similarly, we have
	\begin{equation}
		\begin{split}\label{eq:PWc}
			\hspace{-4mm} P({W_{n,c}} \!\!>\! 0|{W_{n,t}}\! >\! 0,{X_n} \!=\! x,{S_{n,t}} \!=\! s)
			= {e^{ - (\mu_c-\lambda)s}}.
		\end{split}
	\end{equation}
	
	By putting \eqref{eq:Pib} \eqref{eq:Wc>0}-\eqref{eq:Wt=0,Wc>0} \eqref{eq:fW_fenjie2}-\eqref{eq:PWc} together, the proof is completed.

	\section{Proof of Lemma  5}
	Since $W'_{1,n,c}$ and $X_{1,n}$ are independent, we have
	\begin{equation}\label{eq:XW1}
		\mathbb{E}[{X_{1,n}}W'_{1,n,c}|I_n] = \mathbb{E}[{X_{1,n}}|I_n]\mathbb{E}[W'_{1,n,c}|I_n].  \nonumber
	\end{equation}
	
	\subsubsection{$\mathbb{E}[{X_{1,n}}|I_n]$}
	Since $D_{1,n,t}$ and $T_{1,n-1,c}$ are independent, the conditioned PDF of $D_{1,n,t}$ is obtained as
	\begin{equation}
		\begin{split} \label{eq:fDIL}
			f_{D_{1,n,t}|I_n}(y) &= ({\mu _c} - {\lambda _2}){e^{ 	- ({\mu _c} - {\lambda _2})y}}\\
		\end{split}
	\end{equation}

    We have
	\begin{equation}\label{}
		f_{{X_{1,n}}|D_{1,n,t} = y}(t) = {f_{{D_{1,n,t}}|{X_{1,n}} = y}}(t). \nonumber
	\end{equation}
	
	Then it follows that
	\begin{equation}\label{eq:X|Dy}
		\begin{split}
			\hspace{-2mm} \mathbb{E}[X_{1,n}|D_{1,n,t} = y]=&\int_0^{\infty}tf_{{X_{1,n}}|D_{1,n,t} = y}(t)dt\\
			=&\frac{1}{{{\mu _t}}} + y - \frac{1}{{{\mu _t} \!-\! {\lambda _1}}}(1 \!-\! {e^{ - ({\mu _t} - {\lambda _1})y}})
		\end{split}
	\end{equation}
	
	Combining \eqref{eq:fDIL} and \eqref{eq:X|Dy}, we have
	\begin{equation}
		\begin{split} \label{eq:XW1.1}
			\mathbb{E}[X_{1,n}|I_n]=&\int_0^{\infty}\mathbb{E}[X_{1,n}|D_{1,n,t} = y]f_{D_{1,n,t}|I_n}(y)dy\\
			=&\frac{1}{{{\mu _c} - {\lambda _2}}} - \frac{{{\lambda _1}}}{{{\mu _t}({\mu _t} - {\lambda _1})}}\\
			&+\frac{{{\mu _c} - {\lambda _2}}}{{({\mu _t} - {\lambda _1})({\mu _t} + {\mu _c} - \lambda_1-\lambda_2 )}}
		\end{split}
	\end{equation}
	
	\subsubsection{$\mathbb{E}[{W'_{1,n,c}}|I_n]$}
	We evaluate $W'_{1,n,c}$ as
	\begin{equation}\label{}
		W'_{1,n,c} = (T_{1,n-1,c}-D_{1,n,t})^+. \nonumber
	\end{equation}
	
	Given $I_n$, we have
	\begin{equation}\label{}
		\begin{split}
			f_{W'_{1,n,c}|I_n}(t) &=f_{W'_{1,n,c}|W'_{1,n,c}>0}(t)\\ &=(\mu_c-\lambda_1-\lambda_2)e^{-(\mu_c-\lambda_1-\lambda_2)t}.  \nonumber
		\end{split}
	\end{equation}

	\begin{equation}
		\begin{split} \label{eq:XW1.2}
			\mathbb{E}[W'_{1,n,c}|I_n]=\int_0^{\infty}tf_{W'_{1,n,c}|I_n}(t)dt
			=\frac{1}{\mu_c\!-\!\lambda_1\!-\!\lambda_2}
		\end{split}
	\end{equation}
	
     Combining with \eqref{eq:XW1.1}, $\mathbb{E}[{X_{1,n}}{W'_{1,n,c}}|I_n]$ can be obtained.

	\section{proof of Lemma 6}
	We can divide the variables into two groups: $\{\lambda_m\}_{m=1}^M$ and $\{\beta_m,f_m\}_{m=1}^M$.
	Take the second derivative of $\Delta_m$ in equation \eqref{Am} with respect to $\lambda_m$, 
	given $\lambda_{m} > 0$ and $\beta_m$ and $f_m$ are regarded as constants, 
	we have:
		\begin{equation}
				\begin{split}
					\frac{d^2\Delta_m}{d\lambda_m^2}=
					&\frac{2}{(\mu^t_m-\lambda_m)^2} + \frac{2}{(\mu^c_m-\lambda_m)^2} +
					\frac{2}{\lambda_m^3} + \\ &\frac{2(\mu^t_m+\mu_m^c)^2}{\mu^t_m\mu^c_m(\mu^t_m+\mu^c_m-\lambda_m)^2} \ \textgreater 0. \nonumber
				\end{split}
		\end{equation}	
	Therefore,  $\Delta_m$ is a convex function with respect to $\lambda_m$.
	
	Then we fix the value of $\lambda_m$ , 
	and it is easy to observe that $\frac{1}{\mu_m^t}$ and $\frac{1}{\mu_m^c}$ in \eqref{Am} are polynomials.
	As for the first term of \eqref{Am}, i.e., 	$f(\mu^t_m)=\frac{{{(\lambda_m)^2}}}{{(\mu^t_m)^2({\mu^t_m} - \lambda_m )}}$,
	given $\mu_t>0$, 
	we have:
	\begin{equation}
			\begin{split}
				f''(\mu^t_m)= &\frac{6}{(\mu^t_m)^4(\mu^t_m-\lambda_m)} + \frac{4}{(\mu^t_m)^3(\mu^t_m-\lambda_m)^2} \\
				&+ \frac{2}{(\mu^t_m)^2(\mu^t_m-\lambda_m)^3} \ \textgreater 0. \nonumber
			\end{split}
	\end{equation}	
	Therefore,  $f(\Delta_m)$ is a convex function with respect to $\mu^t_m$.
	
	Similiarly, the third term of \eqref{Am} can also be proved to be convex with respect to $\mu^c_m$.
	Moreover, the Hessian matrix of the fifth term is obtained as: 
	\begin{equation}
			H = \frac{1}{\mu^t_m\mu^c_ma^3}
			\left[
			\begin{array}{cc}
				\frac{2}{(\mu_m^t)^2} H_{11} 
				& \frac{1}{\mu_m^t\mu_m^c} H_{12}  \\ 
				\frac{1}{\mu_m^t\mu_m^c} H_{21} 
				& \frac{2}{(\mu_m^c)^2} H_{22}  \nonumber
			\end{array}
			\right],
	\end{equation}
	where $H_{11}=a^2+\mu^t_ma+(\mu_m^t)^2$, $H_{12}=H_{21}=a^2+a(\mu^t_m+\mu^c_m)+2\mu^t_m\mu^c_m$, $H_{22}=a^2+\mu^c_ma+(\mu_m^c)^2$, and $a=\mu^t_m+\mu^c_m-\lambda_m \textgreater 0$.

	It is easy to observe that $H$ is positive-definite:
	\begin{equation}
		\begin{split}
			|H| =& \frac{1}{(\mu_m^t\mu_m^ca)^4}\left[ 3a^2 + 2a(\mu_m^t+\mu_m^c) + 2(\mu_m^t)^2 + \right. \\
			&\left. 2(\mu_m^c)^2 + (\mu_m^t-\mu_m^c)^2 \right] \textgreater 0.\nonumber
		\end{split}	
	\end{equation}  	
	Therefore, the fifth term in \eqref{Am} is also a convex function of $\{\beta_m,f_m\}$.

	To sum up, 
	$\Delta_m$ is a convex function of $\{\beta_m,f_m\}$ given fixed $\lambda_m$.
   Due to the fact that all other constraints are linear, 
   the formulated problem is proved to be a multi-convex problem over its variable sets.

\end{document}